 \documentclass[journal]{IEEEtran}
\usepackage{cite}
\usepackage{multirow}
\usepackage{graphicx}
\usepackage{amssymb,amsmath,times,wasysym}
\usepackage{pb-diagram}
\usepackage{psfrag,balance}
\usepackage{srcltx,verbatim} 

\usepackage{fixltx2e}
\usepackage{multicol}
\usepackage{color}
\usepackage{algorithm,algpseudocode}
\usepackage{algorithm}
\usepackage{enumitem}
\usepackage[table]{xcolor}
\usepackage{amsthm}
\usepackage{subcaption}
\theoremstyle{remark}
\newtheorem*{remark}{Remark}

\newcommand{\norm}[1]{\left\|#1\right\|}

\newcounter{appendx}

\algnewcommand{\LineComment}[1]{\State \(\triangleright\) \textit{#1}}

\newcommand{\real}[1]{\Re\left({#1}\right)}

\renewcommand{\log}[2][]{\mathrm{log}_{#1}\left(#2\right)}

\title{{Tri-Hybrid Cell-Free Massive MIMO Systems} \vspace{-2mm}}

\author{Janith   Dassanayake, \IEEEmembership{Graduate Student Member, IEEE} and  Gayan   Aruma Baduge, \IEEEmembership{Senior Member, IEEE}\vspace{-0mm}
	
\thanks{The authors are with the School of Electrical, Computer and Biomedical Engineering, Southern Illinois University, Carbondale, IL, USA, Email: \{janith.dassanayake, gayan.baduge\}@siu.edu. This work in part has been submitted to the IEEE International Conference on Acoustics, Speech, and Signal Processing (ICASSP) May, 2027 \cite{Dassanayake2026}.} \vspace{-8mm}}

\DeclareMathAlphabet\mathbfcal{OMS}{cmsy}{b}{n}

\begin{document}
\bstctlcite{IEEEexample:BSTcontrol}
\vspace{-6mm}
\maketitle
 
\begin{abstract}

	This paper proposes a tri-hybrid precoder design, cascading digital, analog, and electromagnetic processing layers, for cell-free massive multiple-input multiple-output  (CFmMIMO) systems. We jointly optimize the three-layer precoders and access point (AP) transmit power to maximize global energy efficiency (GEE) under per-AP power and fronthaul constraints. To solve this, we develop an alternating optimization algorithm. The first sub-routine maximizes sum spectral efficiency (SE) for precoder design using weighted minimum mean square error, Riemannian manifold optimization, and gradient ascent over Lorentzian phases. The second optimizes power allocation via Dinkelbach's fractional programming and successive convex approximation. Convergence and computational complexity are also analyzed. Simulation results show the proposed design improves GEE by 87\% and 11\% over fully digital and classical hybrid precoders, respectively, at 20\,dBm transmit power. Moreover, joint power optimization yields 22\% SE and 16\% GEE gains over equal-power baselines, while outperforming co-located architectures in SE. 
    Finally, we identify GEE-optimal regimes for AP and RF chain counts, and demonstrate that increasing dynamic metasurface antenna sizes enhances GEE in low-power regimes with negligible active power overhead.

\end{abstract}

 \vspace{-1mm}



\vspace{-5mm}
\section{Introduction}\label{sec:introduction}
 \vspace{-1mm}

Candidate physical layer technologies for the sixth-generation (6G) and beyond are currently being researched. According to the  international mobile telecommunications (IMT)-2030 framework, established by the international telecommunication union (ITU) in 2023, the peak data rate of 6G is expected to reach 1\,Tbps per user equipment (UE), while 
a minimum of 1\,Gbps UE experienced data rate will be available ubiquitously \cite{Dassanayake2026, Wang2023}. To achieve these data rate targets, 6G physical layer will need to deliver a spectral efficiency (SE) up to 100\,bps/Hz, which translates into 5 times the SE of the fifth-generation (5G) specified in the IMT-2020 framework \cite{Wang2023}.  Although network densification, which leverages spatial frequency reuse,  can significantly enhance the SE  per unit area,  its gains are plateaued when aggregate co-channel interference overwhelms the signal-to-interference-plus-noise ratio (SINR).  The  upper mid-band (7-24\,GHz) or frequency range 3 (FR3)  has recently been proposed for 6G by  ITU to balance the coverage-bandwidth trade-off between the existing sub-6\,GHz (FR1) and millimeter-waves (FR2) \cite{Baduge2026,Bazzi2026}. However,      the aggregate bandwidth in FR3 allocated for 6G is still limited due to many incumbent services.  Hence, a pragmatic 	way  to  linearly
increase the  SE is to boost spatial multiplexing gains by leveraging  large antenna arrays for  multi-user multiple-input multiple-output (MIMO)  \cite{Baduge2026,Bazzi2026}. To this end, a combination of novel antenna architectures, network deployment strategies, and efficient  signal processing techniques must be innovated to boost the SE of  multi-user MIMO envisioned for next-generation wireless systems.

The overall energy consumption of 6G is expected to reduce by 50\,\% compared to the state-of-the-art 5G baseline through architectural innovations in the physical layer, advanced hardware, and intelligent software \cite{Wang2023}. To achieve this target reduction, the energy efficiency (EE) of  the 6G  physical layer must be enhanced by 10 to 100 times compared to 5G \cite{Wang2023}. Specifically, the radio-frequency (RF) chains and their power amplifiers (PAs) of   the base-stations (BS) in the radio access network account for a dominant share of the total energy consumption of mobile networks. Hence, unlike in 5G, the EE has been identified  as a fundamental design criterion for 6G by ITU to lower operational expenses, limit greenhouse gas emissions, and prolong the  battery-life of energy-constrained UE.

Over the past  decades, the development of multi-antenna technology  has undergone many phases. Full analog precoding techniques that  can shape and steer RF beams through phased arrays date back to the early 20th century. They can passively optimize the radiation beampattern  of a  single data stream by using computationally simple, low-cost analog phase-shifters/switches \cite{Pi2011}. However,  full analog precoding provides  merely a logarithmic improvement in the SE through  phased array gain.   
Parallel to advancements in integrated circuits  design and computer processing power in 1990s, full digital precoding became a viable technique to steer multiple data-streams by using  multi-antenna BSs \cite{Massive_MIMO_BOOK_2016}. It requires sophisticated baseband signal processing and an individual  RF chain with a  dedicated digital-to-analog converter (DAC) for each antenna in the array to actively optimize both amplitude and phase of each signal containing   data-streams. It is able to linearly boost the achievable SE by a pre-log factor of $\min(M,N)$, where $M$ is the number of BS antennas and $N$ is the aggregate  of UE antennas \cite{Heath2016}. However,  computational complexity,  implementation cost, and power consumption of full digital precoding are substantially   higher than those of the analog precoding. 
Hence, full digital precoding becomes prohibitively expensive and complicated when the BS antenna array size grows large, specifically for use-cased with FR2,   FR3, and emerging sub-THz  bands. As a remedy, hybrid precoding architectures have recently emerged to balance the trade-offs among the complexity/cost, EE, and SE gains of full digital/analog precoding \cite{Heath2016, Molisch2017}. In hybrid precoding, the number of RF chains can be  reduced  than the BS antenna count by cascading a low-dimensional digital precoder with a high-dimensional analog precoder \cite{Park2017}.    
Depending on how a limited number of RF chains are connected to a large antenna array, hybrid precoding can be classified into several architectures; fully-connected, partially-connected, hybrid partially-fully connected, and dynamic arrays of sub-arrays \cite{Park2017}.

To meet the key performance indicators (KPIs) for   SE and EE, a novel precoding architecture, which distributes precoding across three domains, namely digital, analog, and electromagnetic (EM), has recently emerged. It is referred to as the tri-hybrid precoding. The digital and analog domain processing of tri-hybrid precoder is performed similar to that of the conventional hybrid precoding \cite{Castellanos2025}. In contrast,  EM domain  processing is performed by leveraging reconfigurable antenna surfaces such as dynamic metasurfaces (DMAs) \cite{Shlezinger2021, Castellanos2025}.   To this end, it offloads the sophisticated, energy-intensive spatial processing to  low-power, low-cost PIN diode/varactor-based DMAs.  This architecture enhances  flexibility of using RF chains, phase-shifters/switchers,  and DMAs in designing   multi-user MIMO systems \cite{Shlezinger2021, Castellanos2025}. Thus,  tri-hybrid precoding has  great potential for boosting the achievable EE, while substantially reducing the SE performance  gap  between the full digital and hybrid precoding architectures.  It is also a scalable precoding technique to practically realize   flexible MIMO systems, which are envisioned to leverage  dynamic channel adaptation and advanced spatial processing in the EM domain,  without compromising the achievable EE gains \cite{Castellanos2025}.

Cell-free  wireless architecture has recently been envisioned to overcome several deficiencies of the conventional cell-based counterpart. Specifically, cell-free massive MIMO (CFmMIMO)  aims to  eliminate cell-end UE issues such as severe received signal degradation and co-channel interference by deploying a large number of access points (APs) distributed across a given geographical area \cite{Ngo2017, Ngo2018}.  The  APs coherently serve multiple UEs in the same time-frequency resource element by using  spatial multiplexing to provide a uniform quality-of-service across the entire system \cite{Ngo2017, Ngo2018}.   CFmMIMO has been shown to outperform co-located massive MIMO (mMIMO) by leveraging superior macro-diversity, favorable propagation with reduced path-loss, and mitigating cell-edge issues and inter-cell interference \cite{Ngo2017, Ngo2018}.   The main focus of this work is to explore  EE-optimized tri-hybrid precoder designs for CFmMIMO systems.   To this end, the existing design approaches for conventional hybrid precoding and tri-hybrid precoding  are summarized below.

\vspace{-3mm}

\subsection{Literature Survey}

\subsubsection{Energy-Efficient Hybrid Precoding for Co-located mMIMO}

Hybrid precoding     has matured into a well-established technology for co-located mMIMO systems, with a broad range of design criteria investigated in the literature, spanning hardware complexity reduction, SE/EE maximization, wideband operation, and integration with advanced multiple-access and power transfer schemes in \cite{Yu2016, Liang2014, Park2017, Dai2019} and references therein. Among them, the EE maximization has been one of the important design criteria for  hybrid precoding   \cite{Gao2016, Ribeiro2018, Moghadam2018, Ren2020} and references therein. In \cite{Gao2016}, a successive interference cancellation-based hybrid precoding algorithm is developed for the sub-connected structure, unveiling that the sub-connected architecture requires only one phase shifter per antenna element rather than one per antenna per RF chain. This design  achieves substantially higher EE gains than that by both the fully-connected hybrid and fully-digital precoders under practical power consumption models. In \cite{Ribeiro2018}, the SE-EE trade-off  is characterized under realistic RF hardware loss models. It is shown that insertion losses through power dividers, combiners, and phase-shifters in the fully-connected structure can cause signal-to-noise-ratio degradation. It also reveals that the partially-connected structures with passive phase-shifters offer  EE gains at the cost of weaker SE gains. In \cite{Moghadam2018}, the impact of nonlinear PAs on hybrid precoding is analyzed by using a stochastic distortion model. It also shows that  the spatial directions of the PA-induced distortion are shaped by the precoders and  increasing transmit power beyond a threshold is counter-productive in terms of the achievable EE gains. In \cite{Ren2020}, an RF chain selection framework is  proposed for the sub-connected hybrid precoder, in which  Dinkelbach's algorithm combined with alternating minimization determines the optimal number of active RF chains. This design achieves a better EE-SE  trade-off than that for the fixed RF chain configurations.

\subsubsection{Energy-Efficient Hybrid Precoding for CFmMIMO}
In  \cite{He2022, He2023, Nguyen2022, Guo2023}, design of hybrid precoders have been extended to   CFmMIMO systems. In \cite{He2022}, a dynamic-subarray quantized-phase-shifter hybrid precoder for CFmMIMO is presented. By considering the power consumption in fronthauls
that are required for communications between the APs and the central processing unit (CPU) in cell-free architectures,  the global energy efficiency (GEE) is maximized through a joint design that determines the number of RF chains. This design leverages eigenvalue maximization and solves a power allocation problem by merging fractional programming with non-cooperative game theory and a gradient-assisted binary search algorithm. A wideband CFmMIMO  system is investigated in \cite{He2023}, in which  a beam-squint compensation method based on an iterative heuristic Gram-Schmidt algorithm is proposed for both fully-connected and subarray-based architectures with low-resolution phase-shifters. The per-AP RF chain count is subsequently optimized as a shortest-path problem solved by a bidirectional greedy tree search algorithm to maximize the GEE. In \cite{Nguyen2022}, decentralized and semi-centralized hybrid precoders are designed for uplink mmWave CFmMIMO. It also explores  an adaptive RF chain activation scheme that  selectively turns on RF chains across distributed APs based on channel quality, achieving significant EE gains with only marginal losses in SE. In \cite{Guo2023}, the ergodic EE of a mmWave heterogeneous CFmMIMO system that  employs zero-forcing hybrid precoders is characterized. This design uses a semi-closed-form expression obtained via discretization.  Two heterogeneous gain metrics are also derived to  guide the deployment of heterogeneous APs for CFmMIMO systems.

\subsubsection{Existing Tri-Hybrid precoder Architectures}

	The existing tri-hybrid precoder designs  primarily focus on maximizing the SE. In \cite{Castellanos2025}, a novel tri-hybrid MIMO framework is proposed by cascading  an EM precoding layer into the digital and analog precoding layers in conventional  hybrid architecture. This third EM precoding layer has been realized via a DMA. 
	Specifically, \cite{Castellanos2025} uses a suboptimal sequential   optimization approach to ensure a computationally efficient design  instead of using an optimal joint optimization approach. The achievable SE and EE performance  of the tri-hybrid  precoder of \cite{Castellanos2025} has also been compared against the conventional hybrid precoders. The work in  \cite{Castellanos2025}   reveals that the  tri-hybrid architecture achieves   higher EE across a wide range of input transmit power levels, consistently outperforming fully-analog, fully-digital, and conventional hybrid counterparts, since the  DMA layer     expands the number of radiating elements without a proportional increase in RF chains or phase-shifters.
	
	In \cite{Li2026}, another tri-hybrid precoder is proposed by using  a radiation-center reconfigurable antenna array (RCAA). Its design leverages  a tri-loop alternating optimization (AO) framework to  maximize the SE and a separate fractional-programming-based optimization to maximize  the EE  for a  single-cell co-located MIMO system. It is revealed in  \cite{Li2026}  that the RCAA can achieve SE gains close to that of a fully-digital benchmark while offering EE gains over conventional fixed-position-antenna counterparts. It also shows that the   low-complexity fractional-programming scheme achieves comparable EE performance to a higher-complexity benchmark while reducing execution time by roughly three orders of magnitude, making the design practical for real-time deployment.
	
	 In \cite{Liu2025}, a tri-timescale precoder design framework is proposed, and it updates the EM, analog, and digital precoding layers at progressively faster timescales to reduce pilot overhead and computational complexity.  Again, the achievable SE  is used as the optimization objective in \cite{Liu2025} for its proposed tri-hybrid precoder design. The tri-timescale  design approach in \cite{Liu2025}   reduces the channel estimation overhead and computational complexity relative to a fully real-time joint optimization baseline, since the EM layer is updated only from long-term channel statistics rather than instantaneous CSI, while still achieving SE performance close to that of the real-time benchmark.
	 
	 The tri-hybrid architecture proposed in \cite{Chen2026} uses a pinching-antenna layer as the EM layer, and the pinching  positioning is optimized via the zero-forcing approach. A cross-entropy probabilistic learning framework is also used to maximize the SE. The results in \cite{Chen2026} show that the proposed cross-entropy-based optimization scheme   outperforms a random phase-shifter and pinching-antenna placement baseline, with the performance gain being most pronounced when the number of RF chains does not match the number of waveguides, demonstrating the value of jointly optimizing the phase-shifter network and discrete antenna activation rather than treating them independently.

	Despite recent research developments in tri-hybrid precoder architectures   \cite{Castellanos2025, Li2026, Liu2025, Chen2026}, its  integration  into emerging CFmMIMO systems that require maximization of the GEE remains unexplored. Our motivation and distinct technical contribution with respect to the existing tri-hybrid precoder designs in  open literature \cite{Castellanos2025, Li2026, Liu2025, Chen2026} are elaborated next. 
	

\vspace{-5mm}

\subsection{Motivation}
 \vspace{-1mm}

This work primarily focuses on  designing a  tri-hybrid precoding framework for CFmMIMO systems. To the best of our knowledge, tri-hybrid precoding for CFmMIMO systems  has not been explored in the open literature despite the well-documented SE/EE benefits of cell-free deployments. Moreover, the EE has been recognized as one of the KPIs for 6G, and hence, improving the GEE of CFmMIMO systems is  of particular interest. These facts naturally motivate us to   optimize the tri-hybrid precoders with the explicit design objective  of maximizing the GEE rather than the frequently used SE-based metrics in the open tri-hybrid precoding literature.
With the exception of \cite{Li2026}, none of the other works \cite{Castellanos2025, Liu2025, Chen2026} consider the EE  as a design objective, and even \cite{Li2026} restricts its EE optimization to a co-located architecture. Our optimization procedure also departs from that of \cite{Li2026} in several distinctive ways. In \cite{Li2026}, the EE design optimizes the EM layer through a discrete coordinate-descent search over radiation-center candidates in an outer loop, with the analog beamformer updated via a closed-form approximation in an inner loop, and the power is absorbed directly into the precoder.  In contrast, the proposed GEE-centric tri-hybrid precoder design approach is novel as it jointly updates the EM, analog, and digital precoders through continuous gradient-based, Riemannian manifold, and weighted minimum mean square error (WMMSE) updates within an alternating framework. It is also coupled with a separate Dinkelbach and successive convex approximation (SCA)-based power-allocation stage across distributed APs subject to per-AP power constraints, specific to CFmMIMO systems.

 \vspace{-4mm}
\subsection{Our Distinct Contribution}
 \vspace{-1mm}

This work proposes a  novel  tri-hybrid precoder design   for  orthogonal frequency division multiplexing (OFDM)-based  mmWave CFmMIMO systems. It aims to  maximize the GEE through joint optimization of the three-layer precoders and per-AP power allocation coefficients. Our distinct contribution can be summarized as follows.

\begin{itemize}[leftmargin=*]
	
	\item We consider a CFmMIMO system having multiple  distributed APs, which jointly serve multiple users over  wideband  mmWave channels.
	 Each AP is equipped with a tri-hybrid precoder having  a cascade of  digital, analog, and DMA-based EM precoding layers.
	
	\item We formulate the joint precoder and power allocation design as a GEE maximization problem, capturing the   sum rate across all subcarriers relative to the total power consumption of the CFmMIMO system, including per-AP RF chain, phase-shifter, DMA, and fronthaul power requirements.
	
	\item We propose an AO framework with two sub-routines, in which the precoder and power allocation sub-problems are solved iteratively. Under the adopted power model and a fixed power allocation, the Dinkelbach sub-problem of the precoder stage reduces to sum-SE maximization. The digital, analog, and EM precoders are then updated by the WMMSE method \cite{Shi2011}, Riemannian manifold optimization, and projected gradient ascent, respectively. In the power allocation sub-routine, Dinkelbach's method is combined with a SCA procedure that linearizes the sum SE around the current point and solves a convex quadratic program at each iteration, which can be solved via CVX or Gurobi.
	
	\item We present a comprehensive set of numerical results to demonstrate the performance of the proposed algorithm in terms of both SE and GEE metrics. Our evaluations  include the algorithmic convergence behavior, computational complexity, comparisons against baselines such as fully-digital and conventional hybrid precoding benchmarks, and comparisons between the cell-free and co-located deployments of the tri-hybrid architecture. Design insights are also drawn from our numerical results, revealing the  impact of key system parameters on the achievable SE and GEE.

 \item  Several  key design insights are drawn from our numerical  results. First, our tri-hybrid precoder achieves superior GEE over the fully-digital and classical hybrid precoders owing to its relatively lower overall circuit power consumption. Second, the CF deployment  outperforms the co-located mMIMO in terms of the SE due to the formers macro-diversity benefits. The GEE increases monotonically only up to an optimal AP count,  beyond which further AP densification improves the SE at the expense of the GEE due to the growing aggregate circuit power overhead. Third, there also exists a GEE optimal  RF chain count regime, which does not depend on the  user loading or transmit power allocation. This provides a technically solid hardware dimensioning guideline for   tri-hybrid precoder design architectures. Finally,   the achievable GEE gain can be monotonically increased by increasing the DMA element count at an infinitesimal additional active power cost in the low transmit power regime. These design insights have been validated through a set of rigorous numerical results.


 

\end{itemize}

This work goes beyond our related conference submission \cite{Dassanayake2026}, which optimizes the tri-hybrid precoder under an equal power allocation at each AP. In contrast, this work jointly optimizes the per-AP power allocation, develops an alternating optimization framework, analyzes its computational complexity, and provides a broader performance evaluation/comparison.


  \vspace{-0mm}
  \textbf{Notation:}   The conjugate, transpose, and conjugate
   transpose operations are denoted by $(\cdot)^*$, $(\cdot)^{T}$, and
   $(\cdot)^{H}$, respectively. The $n$th entry and the $\ell_2$-norm of the vector $\mathbf{a}$ are denoted by $[\mathbf{a}]_n$ and $\|\mathbf{a}\|_2$, respectively. 
  The $(m,n)$th  entry,   Frobenius norm, and   trace of the matrix $\mathbf{A}$ are denoted by $[\mathbf{A}]_{mn}$,  $\|\mathbf{A}\|_{F}$, and
  $\mathrm{Tr}(\mathbf{A})$, respectively.   The statistical expectation, real part, Hadamard product, determinant, and ceiling function are denoted by $\mathbb{E}[\cdot]$, $\mathfrak{Re}[\cdot]$, $\odot$, $\det(\cdot)$, and $\lceil\cdot\rceil$,  respectively. 
   The set of complex numbers is denoted by $\mathbb{C}$. The $N \times N$ identity matrix is denoted as $\mathbf{I}_{N}$. 
  The Kronecker delta is denoted by $\delta_{jk}$.

\vspace{-3mm}
\section{System and Signal Models}
 \vspace{-1mm}
\begin{figure}[!t]\centering \vspace{-4mm}
 	\def\svgwidth{210pt} 
 	\fontsize{6}{4}\selectfont 
 	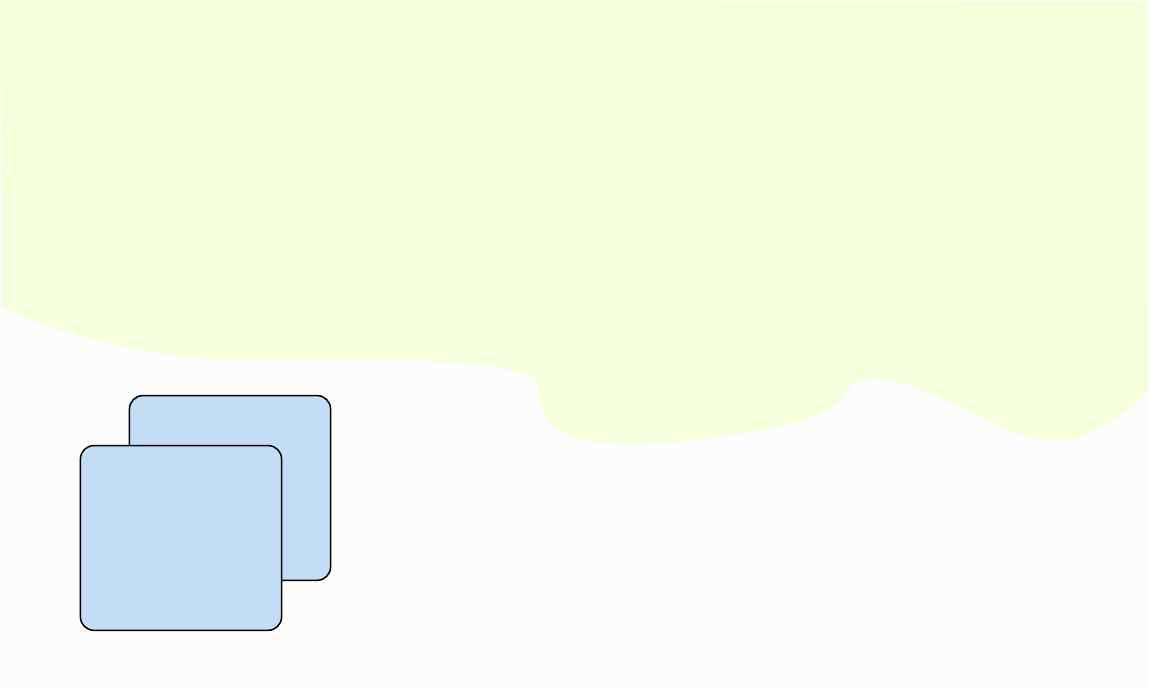 \vspace{-2mm}
 	\caption{Tri-hybrid cell-free massive MIMO}\vspace{-7mm} \label{fig:system_model}
\end{figure}  

\subsection{System Model}\label{sec:system_model}

We consider an OFDM-based  mmWave CFmMIMO   system having $M$ distributed APs, indexed by $m \in \{1, \cdots, M\}$. These APs coherently  serve $K$ UEs each having $N_r$ receive antennas, indexed by $k \in \{1, \cdots, K\}$ in the same time-frequency resource element by using spatial multiplexing. The APs are connected via fronthauls to a CPU. Each AP is equipped with $N_t$ antenna feed ports. Each feed port drives a waveguide-based DMA with $N_{e}$ reconfigurable radiating elements. The total number of radiating elements per AP is given by $N_t^{\mathrm{rad}} = N_t N_{e}$. Each AP has $N_\mathrm{RF}$ RF chains and transmits $N_s$ data-streams per user, satisfying $N_s \leq N_\mathrm{RF} \leq N_t$. The system operates at a carrier frequency $f_0$ with a total bandwidth $B$, divided into $N_c$ orthogonal subcarriers. The frequency of the $n$th subcarrier can be written as 

\vspace{-4mm}
\begin{small}
\begin{equation}\label{eq:fn}
	f_n = f_0 + \frac{B}{N_c}\left(n - 1 - \frac{N_c - 1}{2}\right), \quad n = 1,\ldots,N_c,
\end{equation}
\end{small}
\vspace{-4mm}

\noindent where $\lambda = c/f_0$ is the wavelength at   carrier frequency $f_0$, and $c$ is the speed of light.
The tri-hybrid precoder at the $m$th AP for the $k$th  user in the $n$th subcarrier can be written as 

\vspace{-5mm}
\begin{small}
\begin{eqnarray}\label{eq:tri-hybrid_precoder}
	\mathbf{F}_{mkn} = \mathbf{F}_{\mathrm{EM},m} \mathbf{F}_{\mathrm{RF},m} \mathbf{F}_{\mathrm{BB},mkn},
\end{eqnarray}
\end{small}
\vspace{-5mm}

\noindent where $\mathbf{F}_{\mathrm{EM},m} \in \mathbb{C}^{N_t^{rad} \times N_t}$ is the EM precoder,  $\mathbf{F}_{\mathrm{RF},m} \in  \mathbb{C}^{N_t \times N_\mathrm{RF}}$ is the analog precoder, and $\mathbf{F}_{\mathrm{BB},mkn} \in \mathrm{C}^{N_\mathrm{RF} \times N_s}$ is the baseband precoder. Next, the design aspects of these three precoders are presented.

Firstly, design of the EM precoder can be defined through a  block-diagonal form $\mathrm{blkdiag}(\cdot)$  as 

\vspace{-5mm}
\begin{small}
\begin{eqnarray}\label{eq:block_diag_}
	 \mathbf{F}_{\mathrm{EM},m} = \mathrm{blkdiag} \left( \mathbf{f}_{\mathrm{em},m1},\cdots,\mathbf{f}_{\mathrm{em},mp}, \cdots, \mathbf{f}_{\mathrm{em},mN_t} \right),
\end{eqnarray}
\end{small}
\vspace{-5mm}

\noindent where $\mathbf{f}_{\mathrm{em},mp} \in \mathbb{C}^{N_\mathrm{e} \times 1}$ is the weight vector for the $p$th feed port. The block-diagonal structure arises because each reconfigurable subarray at the DMA is excited by exactly one feed port. Each element of $\mathbf{f}_{\mathrm{em},mp}$ can be written as \cite{Castellanos2025} 

\vspace{-5mm}
\begin{small}
\begin{eqnarray}\label{eq:fem_elem}
	[\mathbf{f}_{\mathrm{em},mp}]_i = \beta_{ip} \alpha_{ip}, \quad i = \{1,\cdots,N_e\},
\end{eqnarray}
\end{small}
\vspace{-5mm}

\noindent where $\beta_{ip} \in (0,1]$ is  the waveguide attenuation coefficient,  and
 $\alpha_{ip}$ is the  Lorentzian-constrained weight. The set of all feasible weights for $\alpha_{i,p}$ is known as the Lorentzian manifold for the DMA, and it can be defined as  \cite{Castellanos2025} 

\vspace{-5mm}
\begin{small}
\begin{eqnarray}\label{eq:lorentz}
	\mathcal{L} = \left\{\alpha_{ip} \in \mathbb{C} : \alpha_{ip} = \tfrac{-j+e^{j\varphi_{ip}}}{2},\; \varphi_{ip} \in [0,2\pi)\right\}.
\end{eqnarray}
\end{small}
\vspace{-5mm}

Secondly, the design of the analog precoder is described as follows:    For the sake of exposition, we consider a fully-connected architecture in the design. 
This choice gives every RF chain access to all feed ports, providing the largest analog beamforming gain for a given $N_\mathrm{RF}$, while the phase-shifter network remains small because the analog layer drives only the $N_t$ feed ports rather than all $N_t^{\mathrm{rad}}$ radiating elements. Partially-connected structures further reduce the phase-shifter count and the power-divider loss, and hence, they can offer a better EE at the cost of a lower SE \cite{Gao2016, Ribeiro2018}. The proposed design can be readily extended to partially-connected architectures by restricting $\mathbf{F}_{\mathrm{RF},m}$ to a block-diagonal support in the manifold update of Section~\ref{sec:GEE_opt} \cite{Park2017}.
In the fully-connected architecture, every RF chain connects to every feed port through a dedicated phase shifter. The underlying  analog precoder $\mathbf{F}_{\mathrm{RF},m} \in \mathbb{C}^{N_t \times N_\mathrm{RF}}$ in \eqref{eq:tri-hybrid_precoder} maps $N_\mathrm{RF}$ RF chains to $N_t$ feed ports, where each entry is a constant-modulus phase-shift in the following form \cite{Yu2016}:
	
	\vspace{-2mm}
    \begin{small}
      	\begin{equation}\label{eq:FRF_structure}
		[\mathbf{F}_{\mathrm{RF},m}]_{ij} = e^{j\theta_{ij}^m}, \quad \text{for}\quad
		\theta_{ij}^m \in [0, 2\pi). 
	\end{equation}  
    \end{small}

	\vspace{-2mm}
	
	\noindent Each element of $\mathbf{F}_{\mathrm{RF},m}$ thus satisfies the constant-modulus constraint as \cite{Yu2016}
	
	\vspace{-5mm}
    \begin{small}
        	\begin{equation}\label{eq:cm}
		\left|[\mathbf{F}_{\mathrm{RF},m}]_{ij}\right|^2 = 1, \quad \forall\; i = 1,\ldots,N_t,\; j = 1,\ldots,N_\mathrm{RF}.
	\end{equation}
    \end{small}

	\vspace{-2mm}
	
	\noindent This analog precoder is shared
	across all $N_c$ subcarriers. The set of all feasible analog precoders
	satisfying \eqref{eq:FRF_structure}-\eqref{eq:cm} is denoted as
	
	\vspace{-5mm}
	\begin{equation}\label{eq:manifold}
		\mathcal{M}_\mathrm{RF} = \left\{\mathbf{F} \in
		\mathbb{C}^{N_t \times N_\mathrm{RF}} :
		|[\mathbf{F}]_{ij}|^2 = 1,\;\forall\,i,j\right\},
	\end{equation}
	\vspace{-5mm}
	
	\noindent which is a smooth Riemannian submanifold of $\mathbb{C}^{N_t \times N_\mathrm{RF}}$ \cite{Yu2016}. 
	
	Thirdly, the design criterion of the digital baseband precoder
	$\mathbf{F}_{\mathrm{BB},mkn} \in \mathbb{C}^{N_\mathrm{RF} \times
		N_s}$ in \eqref{eq:tri-hybrid_precoder} is presented. The digital precoder is 
	user-specific, with each entry taking the  form
	
	\vspace{-4mm}
    \begin{small}
        	\begin{equation}\label{eq:FBB_structure}
		[\mathbf{F}_{\mathrm{BB},mkn}]_{ij} = g_{ij}^{mkn}\,e^{j\phi_{ij}^{mkn}}\!\!,
		\; \text{for} \; g_{ij}^{mkn} \geq 0,	\phi_{ij}^{mkn} \in [0,2\pi),
	\end{equation}
    \end{small}

	\vspace{-2mm}
	
	\noindent where $g_{ij}^{mkn}$ and $\phi_{ij}^{mkn}$ are the amplitude gain
	and phase of the $(i,j)$-th entry for the  $m$th AP, $k$th user, and
	$n$th subcarrier, respectively. It is implemented in the digital domain
	after the DACs and has no structural constraints on either
	$g_{ij}^{mkn}$ or $\phi_{ij}^{mkn}$ beyond the per-AP power constraint. 
	To enable joint tri-hybrid precoder and transmit power optimization, a  power allocation coefficient at the   $m$th AP to the $k$th user is defined as  $\eta_{mk}$. 	
	 To ensure that 	$\eta_{mk}$ serves as  the sole power control variable,
	$\mathbf{F}_{\mathrm{BB},mkn}$ is normalized such that its average
	radiated power gain through the pre-digital precoder is unity.
	Hence, the total precoder power gain $(g_{mk})$, $\forall\,m,k$ can be written
	as

	\vspace{-2mm}
	\begin{small}
		\begin{equation}\label{eq:gmk_def}
			g_{mk} = (1/N_c)\sum_{n=1}^{N_c} \nolimits 
			\left\|\mathbf{F}_{\mathrm{pre},m}
			\mathbf{F}_{\mathrm{BB},mkn}\right\|_F^2 = 1,
		\end{equation}
	\end{small}
	\vspace{-5mm}	
	
	\noindent where $\mathbf{F}_{\mathrm{pre},m} = \mathbf{F}_{\mathrm{EM},m} \mathbf{F}_{\mathrm{RF},m} $.


\vspace{-3mm}
\subsection{Channel Model}\label{sec:channel_model}

The channel between the $m$th AP and the $k$th user is modeled by using the mmWave $D$-delay tap channel model \cite{Alkhateeb2016}. The $d$-th delay tap can be written as

\vspace{-4mm}
\begin{small}
\begin{eqnarray}\label{eq:hd}
	\mathbf{H}_{mk}[d] = \sqrt{(N_t^\mathrm{rad}N_r) / N_{\mathrm{p}}}\sum_{l=1}^{N_{\mathrm{p}}} \nolimits  \gamma_{mkl} p_\mathrm{rc}(dT_s - \tau_{mkl}) \\ \nonumber 
	\times \mathbf{a}_r(\phi_{mkl},f_n) \mathbf{a}_t^H(\theta_{mkl},f_n),
\end{eqnarray}
\end{small}
\vspace{-6mm}

\noindent where $N_{\mathrm{p}}$ and  $\gamma_{mkl}$  are the number of propagation paths and complex path gain of the $l$th path, respectively. Moreover, 
  $p_\mathrm{rc}(dT_s - \tau_{mkl})$, $\phi_{mkl}$ and $\theta_{mkl}$ are the pulse-shaping function, angle of arrival, and angle of departure of the $l$th path, respectively.
The frequency-dependent transmit ($\mathbf{a}_t(y, f_n) $) and receive ($\mathbf{a}_r(y, f_n) $) array steering vectors is defined as 

\vspace{-4mm}
\begin{small}
	\begin{eqnarray}\label{eq:steer}
		\!\!\!\!\!\!\! \mathbf{a}_x(y, f_n) \!=\! \frac{1}{\sqrt{N_x}} \left[1,  e^{j2\pi\frac{f_n}{c}\frac{\lambda}{2}\cos y}, \cdots, e^{j2\pi\frac{f_n}{c}\frac{\lambda}{2}(N_x-1)\cos y}\right]^T \!\!\!\!\!\!,
	\end{eqnarray}
\end{small}
\vspace{-4mm}

\noindent where $y = \phi_{mkl}$ if $x = t$, and  $y = \theta_{mkl}$ if $x = r$. Here, at the carrier frequency $f_n = f_0$, this reduces to the standard half-wavelength-spaced steering vector. 
Then, the frequency-domain channel for the $n$-th subcarrier is obtained by  $N_c$-point discrete Fourier transform of the delay-tap channel as \cite{Alkhateeb2016}

\vspace{-4mm}
\begin{small}
\begin{eqnarray}\label{eq:hfreq}
	\mathbf{H}_{mkn} = \sum_{d=0}^{D-1} \nolimits \mathbf{H}_{mk}[d]\,e^{-j\frac{2\pi n }{N_c}d} \;\in\; \mathbb{C}^{N_r \times N_t^\mathrm{rad}}.
\end{eqnarray}
\end{small}
\vspace{-8mm}

\vspace{-2mm}
\subsection{Signal Model}\label{sec:signal_model}

The data symbol vector intended for the $k$th user on the $n$th subcarrier is denoted by $\mathbf{s}_{kn} \in \mathbb{C}^{N_s}$, which is assumed to be distributed with zero-mean $\mathbb{E}[\mathbf{s}_{kn}]=\mathbf 0$ and unit variance, satisfying  $\mathbb{E}[\mathbf{s}_{kn}\mathbf{s}_{kn}^H] = \mathbf{I}_{N_s}$. The signal transmitted by the $m$th AP on the $n$th subcarrier can be written as

\vspace{-4mm}
\begin{small}
\begin{eqnarray}\label{eq:transmitted_signal}
	\mathbf{x}_{mn} &=& \sum_{k=1}^K \nolimits \sqrt{\eta_{mk}} \mathbf{F}_{mkn}\mathbf{s}_{kn}, 
\end{eqnarray}
\end{small}
\vspace{-5mm}

\noindent where $\eta_{mk}$ denotes the power allocation coefficient from the $m$th AP to the $k$th user. The received signal at the $k$th user on the $n$th subcarrier can be obtained as 

\vspace{-5mm}
\begin{small}
\begin{eqnarray}\label{eq:rx}
	\mathbf{y}_{kn} &=& \sum_{m=1}^M \nolimits  \sqrt{\eta_{mk}}\mathbf{H}_{mkn}\mathbf{F}_{mkn}\mathbf{s}_{kn} \nonumber \\ 
	&+& \sum_{j\neq k}^K \nolimits \sum_{m=1}^M \nolimits \sqrt{\eta_{mj}}  \mathbf{H}_{mkn}\mathbf{F}_{mjn}\mathbf{s}_{jn} + \mathbf{n}_{kn},
\end{eqnarray}
\end{small}
\vspace{-5mm}

\noindent  where  $\mathbf{n}_{kn} \sim \mathcal{CN}(\mathbf{0}, \sigma^2\mathbf{I}_{N_r})$ is an additive white Gaussian noise vector having zero mean and  variance $\sigma^2$ elements. An effective channel as viewed at the front of the digital precoder can  be  defined as $\bar{\mathbf{H}}_{mkn} = \mathbf{H}_{mkn}\mathbf{F}_{\mathrm{EM},m}\mathbf{F}_{\mathrm{RF},m} \in \mathbb{C}^{N_r \times N_\mathrm{RF}}$, and $M$ effective channels pertaining to  all $M$ APs simultaneously serving the  $k$th user on the  $n$th subcarrier	 can be written in a  stacked form as 

\vspace{-5mm}
\begin{small}
\begin{eqnarray}\label{eq:hstack}
	\bar{\mathbf{H}}_{kn} = \left[\bar{\mathbf{H}}_{1kn}, \cdots, \bar{\mathbf{H}}_{mkn}, \cdots, \bar{\mathbf{H}}_{Mkn}\right],
\end{eqnarray}
\end{small}
\vspace{-5mm}

\noindent where $\bar{\mathbf{H}}_{kn} \in \mathbb{C}^{N_r \times MN_\mathrm{RF}} $. Similarly, the digital precoders, weighted by the power allocation coefficients of the corresponding user, can be stacked as

\vspace{-5mm}
\begin{small}
\begin{eqnarray}\label{eq:fbbstack}
	\mathbf{F}_{\mathrm{BB},kn} = \left[\sqrt{\eta_{1k}}\mathbf{F}_{\mathrm{BB},1kn}^T, \cdots, \sqrt{\eta_{Mk}}\mathbf{F}_{\mathrm{BB},Mkn}^T\right]^T,
\end{eqnarray}
\end{small}
\vspace{-5mm}

\noindent where $\mathbf{F}_{\mathrm{BB},kn} \in \mathbb{C}^{MN_\mathrm{RF} \times N_s} $. From \eqref{eq:hstack}-\eqref{eq:fbbstack}, the received signal in \eqref{eq:rx} can be rewritten as

\vspace{-5mm}
\begin{small}
\begin{eqnarray}\label{eq:rx_compact}
	\!\!\!\!\!\!\! \mathbf{y}_{kn} \!=\! \bar{\mathbf{H}}_{kn}\,\mathbf{F}_{\mathrm{BB},kn}\,\mathbf{s}_{kn} \!+\! \sum_{j\neq k}^K \nolimits \bar{\mathbf{H}}_{kn}\,\mathbf{F}_{\mathrm{BB},jn}\,\mathbf{s}_{jn} + \mathbf{n}_{kn},
\end{eqnarray}
\end{small}
\vspace{-5mm}

\noindent where the first term is the desired signal and the second term represents the inter-user interference. 

An achievable sum SE can be written as \cite{Alonzo2019}

\vspace{-5mm}
\begin{small}
\begin{eqnarray}\label{eq:rate}
	\!\!\!\!\!\!\mathcal{R}(\mathbf{F}, \boldsymbol{\eta}) = (1/N_c)\!\! \sum_{k=1}^K \nolimits \sum_{n=1}^{N_c} \nolimits  \mathrm{log}_2 \left|\mathbf{I}_{N_s} + \mathbf{G}_{kn}^H \mathbf{R}_{kn}^{-1}  \mathbf{G}_{kn} \right|\!\!,
\end{eqnarray}
\end{small}
\vspace{-5mm}

\noindent where $\mathbf{G}_{kn} = \bar{\mathbf{H}}_{kn}\,\mathbf{F}_{\mathrm{BB},kn}$, and $\mathbf{R}_{kn}$ is  the  interference-plus-noise covariance matrix observed by the  $k$th user  on the $n$th subcarrier, and it can be defined as

\vspace{-5mm}
\begin{small} 
\begin{eqnarray}\label{eq:Rkn}
    \mathbf{R}_{kn} = \sum_{j\neq k}^K \nolimits \bar{\mathbf{H}}_{kn} \mathbf{F}_{\mathrm{BB},jn} \mathbf{F}_{\mathrm{BB},jn}^H \bar{\mathbf{H}}_{kn}^H + \sigma^2\mathbf{I}_{N_r}.
\end{eqnarray}
\end{small}
\vspace{-5mm}

\noindent In (\ref{eq:rate}), the achievable sum SE is parameterized by the  tri-hybrid precoder ($\mathbf{F}$), which jointly determines the effective spatial multiplexing via tri-hybrid precoding  across all $M$ APs through the three cascaded EM, analog, and digital precoding layers. Moreover, $\mathbf{R}_{kn}$ is also parameterized by     the transmit power allocation vector ($\boldsymbol{\eta}$), which controls the radiated power that each AP allocates to each user.  Hence, $\mathbf{F}$ and $\boldsymbol{\eta}$ serve as the main design components of the proposed tri-hybrid CFmMIMO system. 
On the contrary to the   co-located   systems, in which the transmit power allocation can be absorbed into the precoder magnitudes under a single global power constraint, the cell-free architecture imposes individual per-AP power budgets and hence introduces geographical heterogeneity in AP-user path-losses. This additional design constraint makes    explicit joint optimization of the tri-hybrid precoder $\mathbf{F}$ and the   power allocation  $\boldsymbol{\eta}$  necessary and mutually beneficial for maximizing the GEE of the proposed CFmMIMO system.

\vspace{-3mm}
\subsection{CSI Acquisition and the DMA Model}\label{sec:model_remarks}

 \textbf{\textit{CSI acquisition:}} The main focus of this work is to develop  a GEE optimization framework. Hence, in this work, perfect CSI of $\mathbf{H}_{mkn}$ is assumed to be available at the CPU for all APs, users, and subcarriers, so that the   performance  of the proposed tri-hybrid optimization can be characterized. For wideband OFDM-based mmWave MIMO systems with hybrid precoding, efficient channel estimation techniques have been developed in \cite{Venugopal2017, RodriguezFernandez2018, GaoZ2016}. These techniques exploit the sparse angular-delay structure of mmWave channels to recover the high-dimensional channel from a small number of RF chain observations, via compressive sensing in the time or frequency domain. Similar techniques can be adapted to the proposed tri-hybrid CFmMIMO system, where each AP estimates its local channels from uplink pilots by exploiting the time-division duplex (TDD) channel reciprocity and forwards them to the CPU via fronthaul links \cite{Ngo2017}. The tri-timescale approach in \cite{Liu2025}, in which the EM and analog layers are updated based on channel statistics, can be used to reduce the associated training overhead. However, channel estimation with the DMA-based EM layer, whose Lorentzian-constrained weights restrict the set of realizable training beams, requires a dedicated design.  Hence, it is left for future work.

   \textbf{\textit{DMA model:}} In this work, the Lorentzian-constrained DMA model \eqref{eq:fem_elem}-\eqref{eq:lorentz} is adopted by following \cite{Shlezinger2019, You2023, Shlezinger2021, Castellanos2025}. Here,  $\mathbf{F}_{\mathrm{EM},m}$ is applied to all subcarriers for the sake of simplicity. This approximation is reasonable for the considered system. With $f_0 = 28$\,GHz and $B = 800$\,MHz, all subcarrier frequencies in \eqref{eq:fn} satisfy $|f_n - f_0| \leq B/2$, i.e., they lie within $\pm 1.43\,\%$ of the carrier frequency, and the fractional bandwidth, $B/f_0$, is only about $2.9\,\%$. Over such a narrow band, the DMA response varies only slightly, and hence, the response evaluated at $f_0$ provides a good approximation for all subcarriers \cite{Carlson2026}. However, in practical DMAs, both the Lorentzian response of the radiating elements and the waveguide propagation are frequency selective \cite{Carlson2026}, and this frequency selectivity becomes more pronounced for wider bandwidths and longer waveguides. Incorporating such practical DMA characteristics into the proposed tri-hybrid CFmMIMO design is left for future work.

\vspace{-5mm}
\section{Global Energy Efficiency}\label{sec:power_model}
\vspace{-1mm}
The design criterion of the proposed tri-hybrid precoder and transmit power optimization  is to maximize the  EE subject to system specific constraints. To  this end, a  GEE metric for the proposed tri-hybrid CFmMIMO system can be defined as

  \vspace{-4mm}
  \begin{small} 
  	\begin{eqnarray}\label{eq:gee}
  		\mathcal{G}(\mathbf{F}, \boldsymbol{\eta}) = \mathcal{R}(\mathbf{F}, \boldsymbol{\eta}) / \mathcal P_{\mathrm{T}}(\mathbf{F}, \boldsymbol{\eta}),
  	\end{eqnarray}
  \end{small}
  \vspace{-5mm}
  
  \noindent where $\mathcal{R}(\mathbf{F}, \boldsymbol{\eta})$ is the achievable sum SE, and it is given   in (\ref{eq:rate}). Moreover,  $\mathcal P_{\mathrm{T}}(\mathbf{F}, \boldsymbol{\eta})$  is the total power consumption.

The power consumption model for the proposed CFmMIMO system must be constructed by jointly considering  the component-level framework  for the tri-hybrid precoding,   aggregating power requirements at  each AP, and also accounting for power needed for fronthaul links in cell-free set-up.  
Each RF chain drives $N_t$ feed ports through an $N_t$-way power divider and $N_t$ phase shifters. The EM layer introduces negligible RF insertion loss because the slot coupling is modeled in the EM precoder matrix \cite{Castellanos2025}. The total component loss $L_c$ for the tri-hybrid architecture can be written as 

\vspace{-5mm}
\begin{small} 
\begin{eqnarray}\label{eq:Lth}
	L_c = L_D(N_t) L_\mathrm{PS},
\end{eqnarray}
\end{small}
\vspace{-5mm}

\noindent where $L_D(N_t) = \bar{L}_D^{\lceil\log[2]{N_t}\rceil}$ is the $N_t$-way power divider loss. Here, $\bar{L}_D$ is loss of a two-way power divider. Moreover, $L_\mathrm{PS}$ is the loss due to the phase shifter. Next, the circuit power at the $m$th AP consists of the local oscillator, RF chains, and phase shifters. Each RF chain contains two DAC converters and the RF front-end (mixer, low-pass filter, hybrid coupler), contributing power consumption of  $2P_\mathrm{DAC} + P_\mathrm{RF}$ per chain. Then, the circuit power requirement per AP can be written as

\vspace{-5mm}
\begin{small}
\begin{eqnarray}\label{eq:Pcircth}
	\!\!\!\! P_\mathrm{circ} = P_\mathrm{LO} + N_\mathrm{RF}(2P_\mathrm{DAC} + P_\mathrm{RF}) + N_t\,P_\mathrm{PS} + N_t^{\mathrm{rad}} P_\mathrm{var},
\end{eqnarray}
\end{small}
\vspace{-5mm}

\noindent where $P_\mathrm{var}$ is the bias power of one DMA tuning element. Varactor-tuned elements are reverse biased and draw only a leakage current, and hence, $P_\mathrm{var} \approx 0$ \cite{Shlezinger2021, Castellanos2025}. For PIN-diode-based DMAs, $P_\mathrm{var} > 0$, and the last term in \eqref{eq:Pcircth} grows with $N_e$.
The total power consumed by the proposed tri-hybrid CFmMIMO system can be written as

\vspace{-6mm}
\begin{small} 
\begin{eqnarray}\label{eq:ptotal_with_pc}
	 \mathcal P_{\mathrm{T}}(\mathbf{F}, \boldsymbol{\eta}) =  P_{\mathrm{C}} + c_\mathrm{f} B\mathcal{R}(\mathbf{F}, \boldsymbol{\eta}),
\end{eqnarray}
\end{small}
\vspace{-6mm}

\noindent where $B$ is the bandwidth, $c_\mathrm{f}$ is the traffic-dependent fronthaul power coefficient \cite{He2023}, and $\mathcal{R}(\mathbf{F}, \boldsymbol{\eta})$ is  the achievable SE (\ref{eq:rate}). 
Here, the traffic-dependent fronthaul power is modeled in terms of the aggregate user traffic delivered by the CPU, following \cite{He2023}.
In (\ref{eq:ptotal_with_pc}),  $P_{\mathrm{C}}$ can be written from \eqref{eq:Lth}-\eqref{eq:Pcircth} as 

\vspace{-4mm}
\begin{small} 
\begin{eqnarray}\label{eq:ptotal}
		P_{\mathrm{C}} = (L_c/\zeta_\mathrm{PA}) \sum_{m=1}^M \nolimits  P_{\mathrm{rad},m} + M P_\mathrm{circ} + M P_0  + K p_k,
\end{eqnarray}
\end{small}
\vspace{-5mm}

\noindent where $\zeta_\mathrm{PA}$ is the power amplifier efficiency, $P_0$ is the fixed fronthaul power per AP, and $p_k$ is the per-user circuit power. In (\ref{eq:ptotal}), $P_{\mathrm{rad},m} $ is the  average radiated power from the $m$th AP, and it  can be defined by using \eqref{eq:gmk_def}  as

\vspace{-4mm}
\begin{small} 
\begin{eqnarray}
	\!\!\!\!\! P_{\mathrm{rad},m} = \frac{1}{N_c}\sum_{k=1}^K \nolimits \sum_{n=1}^{N_c}\nolimits \eta_{mk}||\mathbf{F}_{mkn}||_F^2 = \sum_{k=1}^K \nolimits \eta_{mk}. \label{eq:Prad}
\end{eqnarray}
\end{small}
\vspace{-5mm}

\vspace{-3mm}
\section{Global Energy Efficiency Optimization}\label{sec:GEE_opt}
\vspace{-1mm}

This section presents the optimization problem formulation for joint tri-hybrid precoder design and transmit power allocation. Our proposed optimization solution and its underlying optimization  algorithm are also provided.  

\vspace{-3mm}
\subsection{The proposed optimization problem formulation}

The design objective is to maximize GEE (\ref{eq:gee}) by jointly optimizing the tri-hybrid precoder $\mathbf{F}_{mkn}, \,\forall \,m,k,n$ and the transmit power allocation coefficients $\eta_{mk}, \,\forall \,m,k$ subject to a set of system specific constraints. The corresponding optimization problem can be formulated as  
 
\vspace{-5mm}
\begin{small}  
\begin{subequations}
	\begin{eqnarray}
		    \mathbf{P}:  &&\underset{ \mathbf{F}_{\mathrm{EM}}, \mathbf{F}_{\mathrm{RF}}, \mathbf{F}_{\mathrm{BB}}, \boldsymbol{\eta}}{\max}   	\mathcal{G}(\mathbf{F}, \boldsymbol{\eta}) \label{eq:gee_obj} \\
		\text{s.t.}
		&& \mathrm C1:  \mathbf{F}_{\mathrm{RF},m} \in \mathcal{M}_\mathrm{RF}, \quad \forall m = 1,\cdots,M \label{eq:c1} \\
		&& \mathrm C2: [\mathbf{f}_{\mathrm{em},mp}]_i = \beta_{i,p}\alpha_{ip}, \;\forall i,p,m \label{eq:c2} \\
		&& \mathrm C3:  g_{mk} = 1, \quad \forall\,m,k \label{eq:c3} \\
		&& \mathrm C4: \sum_{k=1}^K  \eta_{mk} \leq P_{\max}, \forall m \label{eq:c4} \\
		&& \mathrm C5:  \eta_{mk} \geq 0, \forall m,k,   \label{eq:c5}
	\end{eqnarray}
\end{subequations}
\end{small}
\vspace{-5mm}

\noindent where $	\mathcal{G}(\mathbf{F}, \boldsymbol{\eta})$ is the GEE given in (\ref{eq:gee}), and $P_{\max}$ is the maximum transmit power per AP. 
In (\ref{eq:c1}),   $\mathrm C1$ enforces the constant-modulus constraint of the analog precoder in which  $\mathcal{M}_\mathrm{RF}$ denotes the set of all feasible analog precoders defined as in \eqref{eq:manifold}.
In (\ref{eq:c2}), $\mathrm C2$  imposes the Lorentzian structure of the EM precoder. Moreover,  $\mathrm C3$ in (\ref{eq:c3}) is the precoder normalization, which ensures that the precoder gain through the EM and analog layers is unity, and this is enforced by scaling $\mathbf{F}_\mathrm{BB}$. In  (\ref{eq:c4}), $\mathrm C4$ is the transmit power constraint, and  $\mathrm C5$ in (\ref{eq:c5}) is the non-negativity constraint. 

The optimization problem  $\mathbf{P}$  in (\ref{eq:gee_obj})-(\ref{eq:c5}) is a non-convex fractional program. The numerator of the objective $\mathcal{G}(\mathbf{F}, \boldsymbol{\eta}) = \mathcal{R}(\mathbf{F}, \boldsymbol{\eta}) / \mathcal P_{\mathrm{T}}(\mathbf{F}, \boldsymbol{\eta})$ in (\ref{eq:gee_obj})  involves $\mathrm{log}_2|.|$, and   the denominator  is again coupled to $\mathcal{R}(\mathbf{F}, \boldsymbol{\eta})$ through the fronthaul power requirements as shown in (\ref{eq:ptotal_with_pc}). The constraints $\mathrm C1$  and $\mathrm C2$  define non-convex feasible sets. Hence, to solve $\mathbf{P}$, we resort to an AO procedure. Our optimization solution is described below. 

\vspace{-5mm}
\subsection{The proposed optimization solution}

This subsection presents our proposed AO-based optimization solution to the joint tri-hybrid precoder   and transmit power allocation   problem formulation in (\ref{eq:gee_obj})-(\ref{eq:c5}). It consists of two optimization sub-routines, and they are executed alternately until a convergence is achieved. 

\subsubsection{Sub-routine 1: Fix $\boldsymbol{\eta}$ and optimize $\mathbf{F}$}
First,    the tri-hybrid precoder ($\mathbf{F}$) is optimized while keeping the transmit power allocation coefficients  ($\boldsymbol{\eta}$) fixed. Due to the non-convexity of the corresponding problem,  
a two-layer nested optimization sub-routine is proposed. In the outermost layer, the original objective is transformed into a sequence of parametric sum SE maximization problems by invoking   Dinkelbach's algorithm  \cite{Shen2018} for transforming fractional objectives. Next, three inner layers are formulated to optimize the three precoders ($ \mathbf{F}_{\mathrm{EM}}, \mathbf{F}_{\mathrm{RF}}, \mathbf{F}_{\mathrm{BB}}$) separately. Firstly, in the inner layer 1, the underlying sum SE maximization is   transformed into an equivalent  weighted   sum   mean square error (MSE)  minimization    \cite{Shi2011}  with respect to the digital precoder while keeping  the EM and analog precoders fixed. 
Then, in the inner layer 2, the analog precoder is optimized via manifold optimization. Finally, in the inner layer 3, the EM precoders are updated via projected gradient techniques. The proposed optimization sub-routine 1 is elaborated below.

\begin{itemize}[leftmargin=*] 
	\item \textit{The outermost layer: Dinkelbach's transform and its reduction to sum-SE maximization}
	
\end{itemize}

Dinkelbach's theorem   \cite{Shen2018} is invoked to transform the fractional objective in \eqref{eq:gee_obj}. Applied to $\mathbf{P}$, the following iterative procedure is obtained. At iteration $t$, the parametric subproblem can be written as

\vspace{-5mm}
\begin{small}
\begin{subequations}
	\begin{eqnarray}
	\mathbf{P1}^{(t)}:\quad  \max_{\mathbf{F}_\mathrm{EM}, \mathbf{F}_\mathrm{RF}, \mathbf{F}_\mathrm{BB}} &&  \mathcal{R}(\mathbf{F}) - \nu^{(t)}\,\mathcal P_{\mathrm{T}}(\mathbf{F}) \label{eq:dk_sub} \\
	\text{s.t.} && \mathrm{C1}-\mathrm{C3},
	\end{eqnarray}
\end{subequations}
\end{small}
\vspace{-7mm}

\noindent where the parameter $\nu$ is updated as

\vspace{-3mm}
\begin{small}
\begin{equation}\label{eq:dk_update}
	\nu^{(t+1)} = \mathcal{R}^{(t)}(\mathbf{F}) / \mathcal{P}_{\mathrm{T}}^{(t)}(\mathbf{F}),
\end{equation}
\end{small}
\vspace{-5mm}

\noindent where $\mathcal{R}^{(t)}(\mathbf{F})$ and $\mathcal{P}_{\mathrm{T}}^{(t)}(\mathbf{F})$ are the SE and transmit power evaluated at the solution of $\mathbf{P1}^{(t)}$, respectively.
Here, the dependency of $\boldsymbol{\eta}$ is removed from the notation of  the SE and transmit power to emphasize the fact that $\boldsymbol{\eta}$ is fixed during this sub-routine of the optimization procedure.
The iterations are terminated when $|\mathcal{R}^{(t)}(\mathbf{F}) - \nu^{(t)}\mathcal{P}_{\mathrm{T}}^{(t)}(\mathbf{F})| < \epsilon_\mathrm{DK}$, where $\epsilon_\mathrm{DK}$ is a small positive tolerance. Next, the objective in \eqref{eq:dk_sub} can be rewritten by using \eqref{eq:ptotal_with_pc} as

\vspace{-5mm}
\begin{small}
\begin{equation}\label{eq:dk_simplified}
	\mathcal{R}(\mathbf{F}) - \nu^{(t)}( P_{\mathrm{C}} + c_\mathrm{f} B \mathcal{R}(\mathbf{F})) = \omega^{(t)} \mathcal{R}(\mathbf{F}) - \nu^{(t)}P_{\mathrm{C}},
\end{equation}
\end{small}
\vspace{-5mm}

\noindent where $\omega^{(t)} = 1 - \nu^{(t)} B c_\mathrm{f}$. Under C3, $P_{\mathrm{rad},m} = \sum_{k}\eta_{mk}$ by \eqref{eq:Prad}, and hence, $P_{\mathrm{C}}$ in \eqref{eq:ptotal} does not depend on $\mathbf{F}$ for a fixed $\boldsymbol{\eta}$. Thus, $\nu^{(t)}P_{\mathrm{C}}$ is a constant, and $\omega^{(t)}$ is a constant at the $t$th iteration. Provided that $\omega^{(t)} > 0$, i.e., $\nu^{(t)} < 1/(B c_\mathrm{f})$, the Dinkelbach sub-problem for the inner-loop precoder optimization reduces to a sum SE maximization problem as

\vspace{-5mm}
\begin{small}
\begin{subequations}
	\begin{eqnarray}
		\mathbf{P1}^{(t)}:\quad  \max_{\mathbf{F}_\mathrm{EM}, \mathbf{F}_\mathrm{RF}, \mathbf{F}_\mathrm{BB}} &&  \mathcal{R}(\mathbf{F}) \label{eq:rate_max}  \\
		\text{s.t.} && \mathrm{C1}-\mathrm{C3} \label{eq:rate_C1}.
	\end{eqnarray}
\end{subequations}
\end{small}
\vspace{-6mm}

\noindent The condition $\omega^{(t)} > 0$ is not restrictive in practice. Since $\mathcal{P}_{\mathrm{T}} > c_\mathrm{f}B\mathcal{R}$, every GEE value satisfies $\mathcal{G} < 1/(Bc_\mathrm{f})$, and the Dinkelbach parameter, which is a GEE value by \eqref{eq:dk_update}, therefore always satisfies $\nu^{(t)} < 1/(Bc_\mathrm{f})$. The above problem (\ref{eq:rate_max})-(\ref{eq:rate_C1}) can then be solved through  sum MSE minimization and  manifold-optimization methods at each $t$th iteration.

\begin{remark}
Since the maximizer of $\mathbf{P1}^{(t)}$ does not depend on $\nu^{(t)}$ when $\omega^{(t)} > 0$, repeating the Dinkelbach update in Sub-routine 1 does not change the precoders. The Dinkelbach loop becomes redundant in Sub-routine 1 (see Algorithm~\ref{alg:full}), and the precoder stage maximizes the sum SE for the current power allocation. The GEE objective enters the design through Sub-routine 2, where $\mathcal{P}_{\mathrm{C}}$ depends on $\boldsymbol{\eta}$. We note that this reduction relies on the normalization C3; if C3 were relaxed so that the precoder gain could be traded against the radiated power, the precoder stage would respond to the GEE directly, at the cost of a more involved subproblem.
\end{remark}

\begin{itemize}[leftmargin=*]
	\item \textit{Inner Layer 1: Weighted sum MSE minimization-based Transformation for the Digital Precoder $\mathbf{F}_\mathrm{BB}$}

\end{itemize}

The sum SE optimization sub-problem   $\mathbf{P1}^{(t)}$ in (\ref{eq:rate_max})-(\ref{eq:rate_C1}) can be equivalently transformed into a  sum MSE  minimization problem. Then, the iterative optimization procedure in \cite{Shi2011} can be applied to solve $\mathbf{P1}^{(t)}$ with respect to $\mathbf{F}_\mathrm{BB}$ while keeping $\mathbf{F}_\mathrm{EM}$ and $\mathbf{F}_\mathrm{RF}$ fixed. The WMMSE procedure operates on the stacked precoders $\mathbf{F}_{\mathrm{BB},kn}$ in \eqref{eq:fbbstack}, which include the fixed power allocation coefficients. To start with, two auxiliary variables are defined for each user-subcarrier pair; MMSE receive filter $\mathbf{U}_{kn} \in \mathbb{C}^{N_r \times N_s}$ and $\mathbf{W}_{kn} \succ 0  \in \mathbb{C}^{N_s \times N_s}$. Next, a weighted MSE cost can be defined as

\vspace{-5mm}
\begin{small}
\begin{eqnarray}\label{eq:wmmse_cost}
	\!\!\!\!\!\!\! f(\mathbf{F}_\mathrm{BB},\! \mathbf{U}_{kn},\! \mathbf{W}_{kn}) \!=\! \sum_{k=1}^K\sum_{n=1}^{N_c}\left[\mathrm{Tr}(\mathbf{W}_{kn}\mathbf{E}_{kn}) \!-\! \mathrm{log}|\mathbf{W}_{kn}|\right],
\end{eqnarray}
\end{small}
\vspace{-4mm}

\noindent where $\mathbf{E}_{kn}$ is the MSE matrix, which measures the error between the estimated and true symbols is defined as

\vspace{-4mm}
\begin{small}
\begin{eqnarray}
	\!\!\!\!\!\! &&\mathbf{E}_{kn} = \mathbb{E} \left[(\mathbf{U}_{kn}^H\mathbf{y}_{kn} - \mathbf{s}_{kn})(\mathbf{U}_{kn}^H\mathbf{y}_{kn} - \mathbf{s}_{kn})^H\right] \nonumber \\
	\!\!\!\!\!\! &&= \left(\mathbf{I}_{N_s} - \mathbf{U}_{kn}^H\bar{\mathbf{H}}_{kn}\mathbf{F}_{\mathrm{BB},kn}\right)\left(\mathbf{I}_{N_s} - \mathbf{U}_{kn}^H\bar{\mathbf{H}}_{kn}\mathbf{F}_{\mathrm{BB},kn}\right)^H \nonumber \\
	\!\!\!\!\!\! && + \sum_{j\neq k}^K \nolimits  \mathbf{U}_{kn}^H\bar{\mathbf{H}}_{kn}\mathbf{F}_{\mathrm{BB},jn}\mathbf{F}_{\mathrm{BB},jn}^H\bar{\mathbf{H}}_{kn}^H\mathbf{U}_{kn} + \sigma^2\mathbf{U}_{kn}^H\mathbf{U}_{kn} \label{eq:mse},
\end{eqnarray}
\end{small}
\vspace{-4mm}

\noindent where the first term in the second equality represents the signal distortion, the second term captures the interference leakage through the receive filter, and the third term accounts for the noise enhancement.
The key insight of  the weighted sum MSE minimization framework is that  $ f(\mathbf{F}_\mathrm{BB},\! \mathbf{U}_{kn},\! \mathbf{W}_{kn}) $ achieves the same stationary points as the negative of sum SE when the auxiliary variables $\mathbf{U}_{kn}$ and $\mathbf{W}_{kn}$ are set to their optimal values. Inspired by \cite{Shi2011}, an equivalence can be obtained as

\vspace{-5mm}
\begin{small}
\begin{eqnarray}\label{eq:equivalence}
	\min_{\mathbf{F}_\mathrm{BB}, \mathbf{U}_{kn}, \mathbf{W}_{kn}} f(\mathbf{F}_\mathrm{BB}, \mathbf{U}_{kn}, \mathbf{W}_{kn}) \quad \Longleftrightarrow \quad \max_{\mathbf{F}_\mathrm{BB}} \mathcal{R}(\mathbf{F}_\mathrm{BB}).
\end{eqnarray}
\end{small}
\vspace{-4mm}

\noindent Then, $f(\mathbf{F}_\mathrm{BB}, \mathbf{U}_{kn}, \mathbf{W}_{kn})$ is minimized through the three   steps given below.

\begin{itemize}[leftmargin=0pt, labelwidth=*, align=left]

	\item[\textbf{{Step-(a)}} -] \textit{Receive Filter $\mathbf{U}_{kn}$ Update:}
	To find $\mathbf{U}_{kn}$ that minimizes $\mathrm{Tr}(\mathbf{W}_{kn}\mathbf{E}_{kn})$ in \eqref{eq:wmmse_cost} (or equivalently $\mathrm{Tr}(\mathbf{E}_{kn})$ as $\mathbf{W}_{kn}$ is positive definite), the derivative $\partial\mathrm{Tr}(\mathbf{E}_{kn})/\partial\mathbf{U}_{kn}^*$ is computed and set to zero.
	Then, an optimal MMSE receive filter can be obtained as \cite{Shi2011}

	\vspace{-5mm}
	\begin{small}
		\begin{eqnarray}\label{eq:U_update}
			\mathbf{U}_{kn}^\mathrm{opt} = \mathbf{C}_{kn}^{-1} \bar{\mathbf{H}}_{kn}
			\mathbf{F}_{\mathrm{BB},kn},
		\end{eqnarray}
	\end{small}
	\vspace{-6mm}

	\noindent where $\mathbf{C}_{kn}$ is the effective covariance matrix of the  received signal, defined by

	\vspace{-5mm}
	\begin{small}
		\begin{eqnarray}\label{eq:total_C}
			\mathbf{C}_{kn} = \sum_{j=1}^K  \nolimits \bar{\mathbf{H}}_{kn}\mathbf{F}_{\mathrm{BB},jn}
			\mathbf{F}_{\mathrm{BB},jn}^H\bar{\mathbf{H}}_{kn}^H + \sigma^2\mathbf{I}_{N_r}.
		\end{eqnarray}
	\end{small}
	\vspace{-5mm}

\item[\textbf{Step (b)} -] \textit{Weight Matrix $\mathbf{W}_{kn}$ Update:} Next, the optimal weight matrix is obtained by setting $\partial f / \partial \mathbf{W}_{kn} = 0$ as \cite{Shi2011}

	\vspace{-5mm}
	\begin{small}
		\begin{eqnarray}\label{eq:W_update}
			\mathbf{W}_{kn}^\mathrm{opt} = \mathbf{E}_{kn}^{-1}.
		\end{eqnarray}
	\end{small}
	\vspace{-6mm}

	\item[\textbf{Step (c)} -] \textit{Digital Precoder $\mathbf{F}_{\mathrm{BB},kn}$ Update:} With $\mathbf{U}_{kn}$ and $\mathbf{W}_{kn}$ fixed, the cost~\eqref{eq:wmmse_cost} is quadratic in $\mathbf{F}_{\mathrm{BB},kn}$. By setting the gradient $\partial f / \partial \mathbf{F}_{\mathrm{BB},kn}^* = 0$, and an update for the digital precoder can be written as \cite{Shi2011}

	\vspace{-5mm}
	\begin{small}
		\begin{eqnarray}\label{eq:F_update}
			\tilde{\mathbf{F}}_{\mathrm{BB},kn} = \mathbf{T}_{n}^{-1} \bar{\mathbf{H}}_{kn}^H
			\mathbf{U}_{kn}\mathbf{W}_{kn},
		\end{eqnarray}
	\end{small}
	\vspace{-6mm}

	\noindent where $\mathbf{T}_{n}$ can be defined as

	\vspace{-5mm}
	\begin{small}
		\begin{eqnarray}\label{eq:J_matrix}
			\mathbf{T}_{n} = \sum_{j=1}^K \nolimits \bar{\mathbf{H}}_{jn}^H\mathbf{U}_{jn}\mathbf{W}_{jn}\mathbf{U}_{jn}^H\bar{\mathbf{H}}_{jn} + \mu\,\mathbf{I}_{MN_\mathrm{RF}},
		\end{eqnarray}
	\end{small}
	\vspace{-6mm}

	\noindent where $\mu \geq 0$ is a regularization parameter. Since the precoder normalization C3 is enforced afterwards through \eqref{eq:scale_fbb}, $\mu$ only keeps $\mathbf{T}_n$ well conditioned and bounds the norm of $\tilde{\mathbf{F}}_{\mathrm{BB},kn}$.
	The three steps (a)-(c) are alternately executed  with $I_{W}$ iterations until the relative change in sum rate falls below $\epsilon_{W}$. These steps form a block coordinate descent on the regularized weighted MSE cost, which is therefore non-increasing over the WMMSE iterations \cite{Shi2011}.
	From the resulting digital precoder in \eqref{eq:F_update}, its $m$th block can be extracted and the power allocation coefficient removed as $\tilde{\mathbf{F}}_{\mathrm{BB},mkn} = \tilde{\mathbf{F}}_{\mathrm{BB},kn}[(m-1)N_{\mathrm{RF}} +1:m N_{\mathrm{RF}}, :]/\sqrt{\eta_{mk}}$, for $\eta_{mk} > 0$. Thereby,  the precoder gain for the specific AP-user pair can be computed by using \eqref{eq:gmk_def} as

	\vspace{-2mm}
	\begin{small}
		\begin{equation}\label{eq:gmk_def_tilde}
			\tilde{g}_{mk} = (1/N_c)\sum_{n=1}^{N_c} \nolimits || \mathbf{F}_{\mathrm{pre},m} \tilde{\mathbf{F}}_{\mathrm{BB},mkn} ||_F^2.
		\end{equation}
	\end{small}
	\vspace{-5mm}

	\noindent Then, each per-AP digital precoder is rescaled to ensure the unity-gain constraint C3 as

	\vspace{-5mm}
	\begin{small}
		\begin{eqnarray}\label{eq:scale_fbb}
			\mathbf{F}_{\mathrm{BB},mkn} \leftarrow \tilde{\mathbf{F}}_{\mathrm{BB},mkn}\big/\sqrt{\tilde{g}_{mk}}, \quad \forall\, n,
		\end{eqnarray}
	\end{small}
	\vspace{-5mm}

	\noindent and the stacked precoders are rebuilt through \eqref{eq:fbbstack}.

\end{itemize}

\begin{itemize}[leftmargin=*]
	\item \textit{Inner Layer 2:  Riemannian Manifold Optimization for the Analog Precoder $\mathbf{F}_\mathrm{RF}$}

\end{itemize}

With $\mathbf{F}_\mathrm{EM}$ and $\mathbf{F}_\mathrm{BB}$ fixed, the analog precoder $\mathbf{F}_\mathrm{RF}$ is optimized over the constant-modulus manifold $\mathcal{M}_\mathrm{RF}$.
The constraint $|[\mathbf{F}_{\mathrm{RF},m}]_{ij}|^2 = 1$ defines a smooth manifold, and standard (unconstrained) gradient methods cannot be applied. Instead, inspired by \cite{Absil2008}, the gradient is computed in the ambient Euclidean space and then projected onto the tangent plane of the manifold.

\begin{itemize}[leftmargin=0pt, labelwidth=*, align=left]

		\item[\textbf{Step (i)} -]  \textit{Euclidean Gradient Computation:}
		To compute the Euclidean gradient, the effective channel covariance is formed. To this end,  $\mathbf{Q}_{mkn} = \mathbf{F}_{\mathrm{EM},m}^H\mathbf{H}_{mkn}^H$ is  denoted as the EM-processed channel. Then, the sum channel covariance across all users and subcarriers can be written as

		\vspace{-5mm}
		\begin{small}
			\begin{eqnarray}\label{eq:RQ}
				\mathbf{R}_{\mathbf{Q},m} = (1/N_c)\sum_{k=1}^{K}\nolimits\sum_{n=1}^{N_c} \nolimits \mathbf{Q}_{mkn}\mathbf{Q}_{mkn}^H \;\in\; \mathbb{C}^{N_t \times N_t}.
			\end{eqnarray}
		\end{small}
		\vspace{-5mm}

		\noindent Then, the Euclidean gradient of the channel-power objective with respect to $\mathbf{F}_{\mathrm{RF},m}$ can be written as

		\vspace{-5mm}
		\begin{small}
			\begin{eqnarray}\label{eq:euc_grad}
				\nabla_{\mathbf{F}_{\mathrm{RF},m}} f = 2\,\mathbf{R}_{\mathbf{Q},m}\,\mathbf{F}_{\mathrm{RF},m}.
			\end{eqnarray}
		\end{small}
		\vspace{-5mm}

		\item[\textbf{Step (ii)} -] \textit{Projection to Riemannian Gradient:}
		The tangent space of $\mathcal{M}_\mathrm{RF}$ at point $\mathbf{F}_\mathrm{RF}$ consists of matrices, whose entries are perpendicular (in the complex plane) to the corresponding entries of $\mathbf{F}_\mathrm{RF}$. The Riemannian gradient is obtained by projecting the Euclidean gradient onto this tangent space. The projection can be computed element-wise by removing the radial component as

		\vspace{-5mm}
		\begin{small}
			\begin{eqnarray}\label{eq:riem_grad}
				\mathrm{grad}_\mathcal{M} f = \nabla f - \mathrm{Re}\!\left(\nabla f \odot \mathbf{F}_\mathrm{RF}^*\right) \odot \mathbf{F}_\mathrm{RF},
			\end{eqnarray}
		\end{small}
		\vspace{-6mm}

		\noindent where the term $\mathrm{Re}([\nabla f]_{ij} \cdot [\mathbf{F}_\mathrm{RF}]_{ij}^*)$  in \eqref{eq:riem_grad} computes the real projection of the gradient onto the direction of the current entry, and multiplying by $[\mathbf{F}_\mathrm{RF}]_{ij}$ provides the radial vector to subtract.

		\item[\textbf{Step (iii)} -] \textit{Retraction:}
		A gradient ascent step is taken along the Riemannian gradient, and the result is retracted (projected) back onto the manifold $\mathcal{M}_\mathrm{RF}$ by normalizing each entry to have the required modulus as

		\vspace{-4mm}
		\begin{small}
			\begin{eqnarray}\label{eq:retract}
				[\mathbf{F}_\mathrm{RF}^{(i_{\mathrm{man}}+1)}]_{ij} = \frac{[\mathbf{F}_\mathrm{RF}^{(i_{\mathrm{man}})} + \mu_\mathrm{RF}\,\mathrm{grad}_\mathcal{M} f]_{ij}}{\left|[\mathbf{F}_\mathrm{RF}^{(i_{\mathrm{man}})} + \mu_\mathrm{RF}\,\mathrm{grad}_\mathcal{M} f]_{ij}\right|},
			\end{eqnarray}
		\end{small}
		\vspace{-4mm}

		\noindent where $\mu_\mathrm{RF} > 0$ is the step size and $i_{\mathrm{man}} = \{1, \cdots , I_{\mathrm{man}}\}$, where $I_{\mathrm{man}}$ is the number of iterations in the Inner Layer 2. Here, the division by the absolute value projects each entry onto the unit circle, ensuring that the constant-modulus constraint is satisfied after every iteration.

\end{itemize}

\begin{itemize}[leftmargin=*]
	\item \textit{Inner Layer 3:  Projected Gradient Ascent for the EM Precoder  $\mathbf{F}_\mathrm{EM}$}

\end{itemize}

While keeping $\mathbf{F}_\mathrm{RF}$ and $\mathbf{F}_\mathrm{BB}$ fixed, the EM precoder can be  optimized with respect to the continuous phase parameters $\varphi_{ip} \forall i, p$, at each AP $m$. As the Lorentzian constraint in \eqref{eq:lorentz} is parameterized by $\varphi_{ip}$, the unconstrained gradient ascent can be performed directly in the $\varphi$ domain, and the Lorentzian structure is enforced by  construction.
The derivative of the Lorentzian weight with respect to the phase parameter can be computed as

\vspace{-6mm}
\begin{small}
\begin{eqnarray}\label{eq:dalpha}
	\frac{\partial\alpha_{ip}}{\partial\varphi_{ip}} = \frac{j\,e^{j\varphi_{ip}}}{2}.
\end{eqnarray}
\end{small}
\vspace{-5mm}

To compute the gradient of the sum SE with respect to the EM precoder, the sum SE in \eqref{eq:rate} is rewritten as

\vspace{-4mm}
\begin{small} 
\begin{eqnarray}\label{eq:rate_modified}
	\mathcal{R} = (1/N_c)\sum_{k=1}^K \nolimits \sum_{n=1}^{N_c} \nolimits  r_{kn},
\end{eqnarray}
\end{small}
\vspace{-5mm}

\noindent where $r_{kn} = (1 /\ln 2) \left(\ln|\mathbf{N}_{kn}| - \ln|\mathbf{D}_{kn}|\right)$. Here,  $\mathbf{N}_{kn}$ and $\mathbf{D}_{kn}$ are  the covariance matrices of the total signal-plus-interference-plus-noise  and  interference-plus-noise  at user $k$ on subcarrier $n$, respectively. They   can be written as

\vspace{-4mm}
\begin{small} 
	\begin{eqnarray}
		\mathbf{N}_{kn} &=& \sum_{j=1}^K \nolimits \tilde{\mathbf{C}}_{kjn} \tilde{\mathbf{C}}_{kjn}^H + \sigma^2\mathbf{I}_{N_r}, \label{eq:Nkn} \\
		\mathbf{D}_{kn} &=& \sum_{j\neq k}^K \nolimits  \tilde{\mathbf{C}}_{kjn} \tilde{\mathbf{C}}_{kjn}^H + \sigma^2\mathbf{I}_{N_r}, \label{eq:Dkn}
\end{eqnarray}
\end{small}
\vspace{-4mm}

\noindent where $\tilde{\mathbf{C}}_{kjn} =  \sum_{m=1}^M \sqrt{\eta_{mj}} \mathbf{A}_{mkjn} \in \mathbb{C}^{N_r \times N_s} $ and  $\mathbf{A}_{mkjn} = \mathbf{H}_{mkn}  \mathbf{F}_{\mathrm{EM},m} \mathbf{F}_{\mathrm{RF},m} \mathbf{F}_{\mathrm{BB},mjn}$. Then, the Euclidean gradient of \eqref{eq:rate_modified} with respect to $\mathbf{F}_{\mathrm{EM},m}^*$ can be obtained via $\partial \ln|\mathbf{X}| = \mathrm{Tr}(\mathbf{X}^{-1}\partial\mathbf{X})$ as 

\vspace{-4mm}
\begin{small}
\begin{eqnarray}\label{eq:em_euc_grad}
	\!\!\!\!\frac{\partial \mathcal{R}}{\partial \mathbf{F}_{\mathrm{EM},m}^*} &=& \frac{1}{N_c \ln 2}\sum_{k=1}^{K}\sum_{j=1}^{K}\sum_{n=1}^{N_c} \sqrt{\eta_{mj}}\,\mathbf{H}_{mkn}^H \nonumber \\
	&& \!\!\!\!\!\!\!\!\!\!\!\!\!\!\!\!\!\!\!\! \times \left(\mathbf{N}_{kn}^{-1} - (1-\delta_{jk})\mathbf{D}_{kn}^{-1}\right) \tilde{\mathbf{C}}_{kjn} \left(\mathbf{F}_{\mathrm{RF},m}\mathbf{F}_{\mathrm{BB},mjn}\right)^H\!.
\end{eqnarray}
\end{small}
\vspace{-4mm}

\noindent Only the entries of \eqref{eq:em_euc_grad} on the block-diagonal support of $\mathbf{F}_{\mathrm{EM},m}$ are needed, and all $N_t N_e$ phase parameters of an AP share the same Euclidean gradient. Then, the gradient of the sum SE with respect to $\varphi_{ip}$ can be derived  by using the chain-rule as

\vspace{-4mm}
\begin{small}
\begin{eqnarray}\label{eq:em_grad}
	\frac{\partial \mathcal{R}}{\partial\varphi_{ip}} = 2\,\mathrm{Re}\left\{\left[\frac{\partial \mathcal{R}}{\partial \mathbf{F}_{\mathrm{EM},m}^*}\right]_{i'p}^{*} \beta_{ip} \frac{j\,e^{j\varphi_{ip}}}{2}\right\},
\end{eqnarray}
\end{small}
\vspace{-3mm}

\noindent where $i' = (p-1)N_e + i$ is the global row index in $\mathbf{F}_\mathrm{EM}$ corresponding to the $i$-th element of the $p$-th feed port.
Then, gradient ascent step is performed on each phase parameter as

\vspace{-4mm}
\begin{small}
\begin{eqnarray}\label{eq:phi_update}
		\varphi_{ip}^{(i_\mathrm{EM}+1)} = \left(\varphi_{ip}^{(i_\mathrm{EM})} + \mu_\mathrm{EM}\,\frac{\partial \mathcal R }{\partial\varphi_{ip}}\right) \bmod 2\pi,
\end{eqnarray}
\end{small}
\vspace{-4mm}

\noindent where $\mu_\mathrm{EM} > 0$ is the step size, $i_\mathrm{EM} \in \{1,\ldots,I_\mathrm{EM}\}$ is the iteration index, and the modulo operation ensures that the phase remains in $[0, 2\pi)$.
After all phases are updated, each Lorentzian weight is recomputed as $\alpha_{ip} = (-j + e^{j\varphi_{ip}})/2$. Then, each subarray vector can be formed as $[\mathbf{f}_{\mathrm{em},mp}]_i = \beta_{ip}\,\alpha_{ip}$, and the block-diagonal EM precoder can be obtained via \eqref{eq:block_diag_}. The above construction satisfies the Lorentzian constraint at every iteration.

\subsubsection{Sub-routine 2:  Fix $\mathbf{F}$ and optimize $\boldsymbol{\eta}$}

In this second sub-routine, while keeping all precoders fixed, the optimization problem $\mathbf{P}$ is solved with respect to $\boldsymbol{\eta}$. To this end, the sum SE in \eqref{eq:rate_modified} is written as a function of $\boldsymbol{\eta}$, denoted by $\mathcal{R}(\boldsymbol{\eta})$, where $\mathbf{N}_{kn}$ and $\mathbf{D}_{kn}$ in \eqref{eq:Nkn}-\eqref{eq:Dkn} depend on $\boldsymbol{\eta}$ through $\tilde{\mathbf{C}}_{kjn} = \sum_{m=1}^M \sqrt{\eta_{mj}} \mathbf{A}_{mkjn}$.
When the  precoders are fixed, the quantities $\mathbf{A}_{mkjn}$, $g_{mk}$ in \eqref{eq:gmk_def}, and effective channels in \eqref{eq:hstack} are constants. Then, the Dinkelbach sub-problem with respect to $\boldsymbol{\eta}$ can be formulated via \eqref{eq:dk_simplified} as 

\vspace{-4mm}
\begin{small} 
	\begin{subequations}
		\begin{eqnarray}
			\mathbf{P2}^{(t),(l)}:\quad  \max_{\boldsymbol{\eta}} &&  \omega^{(t)} \mathcal{R}(\boldsymbol{\eta}) - \kappa^{(t)}\sum_{m= 1}^M \nolimits \sum_{k = 1}^K \nolimits \eta_{mk} \label{eq:P2_obj}\\
			\text{s.t.} &&  \, \mathrm{C4}-\mathrm{C5}, \label{eq:P2_C1}
		\end{eqnarray}
	\end{subequations}
\end{small}
\vspace{-6mm}

\noindent where $\kappa^{(t)} = \nu^{(t)} L_c/\zeta_\mathrm{PA}$, and $\omega^{(t)} > 0$ holds for the same reason as in Sub-routine 1.  As $\mathcal{R}(\boldsymbol{\eta})$ is non-convex in $\boldsymbol{\eta}$, we resort to an SCA procedure to solve $\mathbf{P2}^{(t)}$. At each SCA iteration $\ell =\{ 0,  \cdots, N_\mathrm{SCA}\}$ within Dinkelbach iteration $t = \{0, \cdots, I_{\mathrm{DK}}\}$, the sum SE can be linearized around $\boldsymbol{\eta}$ via first-order Taylor expansion as 

\vspace{-4mm}
\begin{small} 
\begin{eqnarray}\label{eq:taylor}
	\!\!\!\!\!\!\! \mathcal{R}(\boldsymbol{\eta}) \approx \mathcal{R}(\boldsymbol{\eta}^{(l)}) + \sum_{m = 1}^M \sum_{k=1}^K  \rho_{mk}^{(l)}  \left(\eta_{mk} - \eta_{mk}^{(l)}\right) =  \tilde{\mathcal{R}}(\boldsymbol{\eta}, \boldsymbol{\eta}^{(l)}) ,
\end{eqnarray}
\end{small}
\vspace{-3mm}

\noindent where $\rho_{m,k} = \frac{\partial \mathcal{R} (\boldsymbol{\eta}) }{\partial\eta_{mk}}$ is the gradient of  $\mathcal{R}(\boldsymbol{\eta})$ with respect to $\eta_{m,k}$ and can be found as 

\vspace{-5mm}
\begin{small} 
\begin{eqnarray}\label{eq:grad_R}
				\rho_{m,k} = \frac{1}{N_c}\sum_{n=1}^{N_c} \left[\frac{\partial r_{kn}}{\partial\eta_{mk}} + \sum_{k'\neq k}^K \frac{\partial r_{k'n}}{\partial\eta_{mk}} \right],
\end{eqnarray}
\end{small}
\vspace{-3mm}

\noindent where the derivatives in \eqref{eq:grad_R} can be derived as 

\vspace{-4mm}
\begin{small} 
\begin{eqnarray}\label{eq:case1_simple}
	\frac{\partial r_{kn}}{\partial\eta_{mk}} &=& \frac{\eta_{mk}^{-\frac{1}{2}}}{\ln 2}  \mathfrak{Re}  \left[\mathrm{Tr} \left( \tilde{\mathbf{C}}_{kkn}^H \mathbf{N}_{kn}^{-1} \mathbf{A}_{mkkn}\right)\right], \\ 
	\frac{\partial r_{k'n}}{\partial\eta_{mk}} &=& \frac{\eta_{mk}^{-\frac{1}{2}}}{\ln 2} \mathfrak{Re}  \left[\mathrm{Tr} \left(\tilde{\mathbf{C}}_{k'kn}^H \left(\mathbf{N}_{k'n}^{-1} - \mathbf{D}_{k'n}^{-1}\right)\mathbf{A}_{mk'kn}\right)\right]. \label{eq:case2_simple}
\end{eqnarray}
\end{small}
\vspace{-4mm}

Then, the optimization problem for each Dinkelbach iteration $t$ can be formulated as 

\vspace{-5mm}
\begin{small} 
	\begin{subequations}
		\begin{eqnarray}
			\!\!\!\!\!\!\!\!\!\!\! \mathbf{P2}^{(t)}\!\!:\!    \max_{\boldsymbol{\eta}} &&  \omega^{(t)} \tilde{\mathcal{R}}(\!\boldsymbol{\eta}, \boldsymbol{\eta}^{(l)}\!) \!-\! \kappa^{(t)}\!\!\sum_{m= 1}^M  \! \sum_{k = 1}^K \!  \eta_{mk}  \!-\! \frac{1}{2\tau}||\boldsymbol{\eta}\!-\!\boldsymbol{\eta}^{(l)}||^2 \label{eq:P2_obj1}\\
			\text{s.t.} &&  \, \mathrm{C4}-\mathrm{C5}, \label{eq:P2_C1_1}
		\end{eqnarray}
	\end{subequations}
\end{small}
\vspace{-6mm} 

\noindent where the third term in the objective in \eqref{eq:P2_obj1} is a proximal regularizer with parameter $\tau > 0$, which limits the step size and avoids oscillation between SCA iterations. Here, the above sub-problem is a convex quadratic program, and it can be  directly solved via efficient convex solvers such as Gurobi, MOSEK or Matlab-based CVX \cite{Boyd2004}.

\textbf{\textit{Algorithmic Convergence:}} The proposed algorithm retains the best solution found across the alternating iterations, and hence, the achieved GEE is non-decreasing over the iterations. The GEE is upper bounded as $\mathcal{R}$ is finite for $\eta_{mk} \leq P_{\max}$ and $\mathcal{P}_{\mathrm{T}} \geq M(P_\mathrm{circ} + P_0) + Kp_k > 0$. Thus, the algorithmic convergence  in terms of the GEE is guaranteed. Each sub-routine further improves the solution efficiently. The WMMSE updates monotonically decrease the weighted MSE cost \cite{Shi2011}. The manifold and Lorentzian phase updates satisfy the constant-modulus Lorentzian constraints at every step. The proximal term in \eqref{eq:P2_obj1} stabilizes the SCA iterations. Consequently, the proposed algorithm converges within only a few alternating iterations as   verified in Section~\ref{sec:numerical}.

\vspace{-4mm}
\section{Computational Complexity Analysis}\label{sec:complexity}

In this section, the computational complexity of the proposed AO-based algorithm is analyzed. The  weighted sum MSE minimization iterations,  manifold gradient steps, and  EM gradient steps are defined as $I_W$, $I_\mathrm{man}$, and $I_\mathrm{EM}$, respectively. In sub-routine 1, the computational complexities for digital precoder update ($C_{\mathbf{F}_\mathrm{BB}}$), manifold optimization ($C_\mathrm{man}$), and projected gradient step ($C_\mathrm{EM}$), respectively, can be determined  as 

\vspace{-6mm}
\begin{small} 
	\begin{subequations}
		\begin{eqnarray}
			C_{\mathbf{F}_\mathrm{BB}} &=& \mathcal{O} \left(I_W N_c \left(K(MN_\mathrm{RF})^2 N_s + (MN_\mathrm{RF})^3\right) \right),  \\ 
			C_\mathrm{man} &=& \mathcal{O} \left( M K N_c N_r (N_t^{\mathrm{rad}} + N_t^2) + I_\mathrm{man} M N_t^2 N_\mathrm{RF} \right), \\
			C_\mathrm{EM} &=& \mathcal{O} \big( I_\mathrm{EM} \big[ M K^2 N_c N_r (N_t^{\mathrm{rad}} + N_r N_s + N_s N_t + N_\mathrm{RF} N_s) \nonumber \\
			&& +\, K N_c N_r^3 \big] \big).
		\end{eqnarray}
	\end{subequations}
\end{small}
\vspace{-6mm}

\noindent In $C_\mathrm{man}$, the first term accounts for forming $\mathbf{R}_{\mathbf{Q},m}$ in \eqref{eq:RQ}, which exploits the block-diagonal structure of $\mathbf{F}_{\mathrm{EM},m}$. In $C_\mathrm{EM}$, the first term accounts for forming $\tilde{\mathbf{C}}_{kjn}$ and the block-diagonal entries of \eqref{eq:em_euc_grad}. The second term tackles   the inverses of $\mathbf{N}_{kn}$ and $\mathbf{D}_{kn}$. Since all $N_t N_e$ phase parameters of an AP share the Euclidean gradient in \eqref{eq:em_euc_grad}, the gradient is evaluated once per EM iteration rather than once per phase parameter.

\noindent The   computational  complexity per alternating iteration in sub-routine 1 can be written as $C_{\mathrm{B}1} = C_{\mathbf{F}_\mathrm{BB}} +	C_\mathrm{man} + C_\mathrm{EM}$.

\noindent For sub-routine 2, the computational complexities for pre-computation ($C_\mathrm{pre}$), matrix inversions ($C_\mathrm{inv}$), SCA gradient calculations ($C_\mathrm{grad}$), and CVX quadratic program ($C_\mathrm{CVX}$), respectively, can be computed  as 

\vspace{-5mm}
\begin{small} 
	\begin{subequations}
		\begin{eqnarray}
			C_\mathrm{pre} &=& \mathcal{O}\!\left(M K^2 N_c
			(N_r N_t^\mathrm{rad} N_\mathrm{RF} + N_r N_\mathrm{RF}
			N_s)\right), \\
			C_\mathrm{inv} &=& \mathcal{O}\big(K N_c (K N_r^2 N_s + N_r^3)\big), \\ 
			C_\mathrm{grad}  &=& \mathcal{O} \left(K^2 N_c
			(N_r^2 N_s + M N_r N_s^2)\right), \\
			C_\mathrm{CVX} &=& \mathcal{O} \left((MK)^{3.5}\log[]{1/\epsilon}\right),			 
		\end{eqnarray}
	\end{subequations}
\end{small}
\vspace{-7mm}

\noindent where $C_\mathrm{CVX}$ is the cost of a primal-dual interior-point method for a convex quadratic program with $MK$ variables and accuracy $\epsilon$ \cite{Boyd2004}. In sub-routine 2, the Dinkelbach loop runs $N_\mathrm{DK}$ times, and
within each Dinkelbach iteration the SCA runs $N_\mathrm{SCA}$
times. Then, the   computational complexity of  sub-routine 2  can be written as $C_{\mathrm{B}2} = C_\mathrm{pre} + N_\mathrm{DK} N_\mathrm{SCA} ( C_\mathrm{inv} + C_\mathrm{grad} + C_\mathrm{CVX} )$.
\noindent Finally, the overall computational  complexity of the proposed optimization algorithm over $N_\mathrm{alt}$ alternating iterations can be written as $C_\mathrm{total} = N_\mathrm{alt}\!\left(C_{\mathrm{B}1} + C_{\mathrm{B}2}	\right)$.


\begin{algorithm}
	\caption{Joint Global Energy Efficiency Maximization for Tri-Hybrid CFmMIMO system}
	\label{alg:full}
	\begin{algorithmic}[1]
		\Require Channels $\{\mathbf{H}_{mkn}\}$, $P_{\max}$,
		$L_c$, $\zeta_\mathrm{PA}$, $\tau$,
		$I_W$, $I_\mathrm{man}$, $I_\mathrm{EM}$,
		$N_\mathrm{DK}$, $N_\mathrm{SCA}$, $N_\mathrm{alt}$
		\Ensure Optimized $\mathbf{F}_\mathrm{EM}^*$,
		$\mathbf{F}_\mathrm{RF}^*$, $\mathbf{F}_\mathrm{BB}^*$,
		$\boldsymbol{\eta}^*$, $\mathrm{GEE}^*$
		
		\State \textbf{Initialize:} $\eta_{m,k} \leftarrow
		P_{\max}/K$, $\forall\,m,k$; \;
		$\mathbf{F}_{\mathrm{RF},m} \leftarrow $ random unit-modulus,
		$\forall\,m$; \;
		$\varphi_{i,p} \leftarrow$ random $\in[0,2\pi)$,
		$\forall\,i,p,m$; form $\mathbf{F}_{\mathrm{EM},m}$ via \eqref{eq:block_diag_} and $\mathbf{F}_{\mathrm{pre},m} \leftarrow \mathbf{F}_{\mathrm{EM},m}\mathbf{F}_{\mathrm{RF},m}$, $\forall\,m$
		
		\For {$t_\mathrm{alt} = 1, 2, \ldots, N_\mathrm{alt}$}
		
		\State \textbf{$\triangleright$ Sub-routine 1: Fix $\boldsymbol{\eta}$, optimize $\mathbf{F}_\mathrm{BB}$, $\mathbf{F}_\mathrm{RF}$, $\mathbf{F}_\mathrm{EM}$}
		
		\State Build $\bar{\mathbf{H}}_{kn}$ via $\mathbf{F}_{\mathrm{pre},m}$ and $\mathbf{F}_{\mathrm{BB},kn} \forall\,k,n$
		\State Initialize $\mathbf{F}_{\mathrm{BB},kn} \forall\,k,n$
		\For{$i_W = 1, \ldots, I_W$}
		\State Update $\mathbf{U}_{k,n}$ via \eqref{eq:U_update},
		$\forall\,k,n$
		\State Compute $\mathbf{E}_{k,n}$ and update
		$\mathbf{W}_{k,n} \leftarrow
		\mathbf{E}_{k,n}^{-1}$,
		$\forall\,k,n$
		\State Update $\mathbf{F}_{\mathrm{BB},kn}$ via
		\eqref{eq:F_update}, $\forall\,k,n$
		\If{$ |\mathcal{R}^{(i_W)} \!-\! \mathcal{R}^{(i_W - 1)} |/|\mathcal{R}^{(i_W - 1)} | \!<\! \epsilon_W$}
		\textbf{break}
		\EndIf
		\EndFor
		\State Normalize $\mathbf{F}^{\mathrm{opt}}_{\mathrm{BB},kn}$ via \eqref{eq:scale_fbb}
		
		\For{each AP $m$}
		\State Compute $\mathbf{R}_{\mathbf{Q},m}$
		via channel and $\mathbf{F}_{\mathrm{EM},m}$
		\For{$i_\mathrm{man} = 1, \ldots, I_\mathrm{man}$}
		\State Compute $\nabla_{\mathbf{F}_\mathrm{RF}} f$ via \eqref{eq:euc_grad}, Compute $\mathrm{grad}_\mathcal{M} f$ via \eqref{eq:riem_grad}, Compute $[\mathbf{F}_{\mathrm{RF},m}^{(i_\mathrm{man}+1)}]_{ij}$ via  \eqref{eq:retract} 
		\EndFor
		\EndFor
		
		\For{$i_\mathrm{EM} = 1, \ldots, I_\mathrm{EM}$}
		\For{each AP $m$, feed port $p$, element $i$}
		\State Compute $\partial \mathcal R/\partial\varphi_{ip}$
		via \eqref{eq:em_euc_grad}-\eqref{eq:em_grad},  Compute $\varphi^{(i_\mathrm{EM}+1)}_{ip}$ via \eqref{eq:phi_update}
		\EndFor
		\State Recompute $\alpha_{ip}$ and 
		update $\mathbf{F}_{\mathrm{EM},m}$
		\EndFor
		
		\State Update $\mathbf{F}_{\mathrm{pre},m} \leftarrow
		\mathbf{F}_{\mathrm{EM},m}\mathbf{F}_{\mathrm{RF},m}$, $\forall\,m$
		\State Rescale $\mathbf{F}_{\mathrm{BB},mkn}$ with the updated $\mathbf{F}_{\mathrm{pre},m}$
		
		\State \textbf{$\triangleright$ Sub-routine 2: Fix $\mathbf{F}$, optimize $\boldsymbol{\eta}$}
		
		\State $\mathbf{A}_{mkjn} \leftarrow
		\mathbf{H}_{mkn}\mathbf{F}_{\mathrm{pre},m}
		\mathbf{F}_{\mathrm{BB},mjn}$,
		$\forall\,m,k,j,n$
		
		\State Initialize $\boldsymbol{\eta} \leftarrow
		\boldsymbol{\eta}^\mathrm{init}$; \quad
		$\nu \leftarrow 0$
		
		\For{$t = 0, 1, \ldots, N_\mathrm{DK}$}
		\State Compute $\kappa \leftarrow \nu L_c/\zeta_\mathrm{PA}$ and
			\quad
			$\omega \leftarrow 1 - \nu B c_\mathrm{f}$
		
		\For{$\ell = 0, 1, \ldots, N_\mathrm{SCA}$}
		\State Compute $\tilde{\mathbf{C}}_{kjn} \leftarrow
		\sum_m\sqrt{\eta_{mj}}	\mathbf{A}_{mkjn}$, $\forall\,k,j,n$
		\State Compute $\mathbf{N}_{kn}$,
		$\mathbf{D}_{kn}$,
		$\mathbf{N}_{kn}^{-1}$,
		$\mathbf{D}_{kn}^{-1}$, $\forall\,k,n$
		\State Compute gradient
		$\rho_{m,k}$ via \eqref{eq:grad_R}-\eqref{eq:case2_simple},
		$\forall\,m,k$
		\State Solve $(\mathbf{P2}^{(\ell)})$ (\eqref{eq:P2_obj1}-\eqref{eq:P2_C1_1}) via CVX
		
		\If{$\norm{\boldsymbol{\eta}^{(\ell+1)}
				-\boldsymbol{\eta}^{(\ell)}}
			< \epsilon_\mathrm{SCA}$}
		\textbf{break}
		\EndIf
		\EndFor
		
		\State Evaluate $R(\boldsymbol{\eta})$ and
			$P_T(\boldsymbol{\eta})$
			\State $\nu \leftarrow
			R(\boldsymbol{\eta})/P_T(\boldsymbol{\eta})$
			\If{$|R - \nu P_T| < \epsilon_\mathrm{DK}$}
		\textbf{break}
		\EndIf
		\EndFor
		
		\State Track best: if $\nu > \mathrm{GEE}^*$,
			update $\mathrm{GEE}^* \leftarrow \nu$,
		$\boldsymbol{\eta}^* \leftarrow \boldsymbol{\eta}$,
		$\mathbf{F}^* \leftarrow \mathbf{F}$
		\EndFor
	\end{algorithmic}
\end{algorithm}

\vspace{-2mm}
\section{Numerical Results}\label{sec:numerical}
\vspace{-0mm}

A carrier frequency of $f_0 = 28$\,GHz with a bandwidth of $B = 800$\,MHz is used for the   wideband OFDM-based mmWave CFmMIMO system. It consists of $M = 8$ APs and $K = 3$ UEs, all deployed uniformly at random within a $100\,\text{m} \times 100\,\text{m}$ service area. Each AP is equipped with $N_t = 8$ antenna feed ports, $N_\mathrm{RF} = 4$ RF chains, and $N_e = 4$ reconfigurable radiating elements per feed port, yielding a total of $N_t^\mathrm{rad} = 32$ radiating elements per AP. Each UE has $N_r = 4$ receive antennas and receives $N_s = 2$ independent data streams. The OFDM waveform uses $N_c = 64$ subcarriers. 
The wideband channel between each AP-UE pair is modeled via \eqref{eq:hd}-\eqref{eq:hfreq} with $N_p = 5$ propagation paths and $D = 5$ delay taps. The path-loss follows the 3GPP Urban Micro (UMi) model \cite{3GPP2024}, given by $\mathrm{PL}(\text{dB}) = 32.4 + 20\log[10]{f_0} + 30\log[10]{d}$, where $d$ denotes the AP-UE distance in meters. 
The hardware power consumption parameters are adopted from \cite{Castellanos2025} and \cite{Ribeiro2018}. They  are furnished in Table \ref{tab:power_params}. The EM precoding layer employs a waveguide attenuation model with per-element decay constant of 0.05 with  $\beta_{i,p} = e^{-0.05(i-1)}$ for $i = 1, \ldots, N_e$. The varactor diode power consumption is assumed to be negligible ($P_\mathrm{var} \approx 0$).

\textbf{\textit{Baselines:}} All baselines use the same channels, the   power allocation sub-routine, and the   power model with architecture-specific terms. The fully-digital AP has $N_t^{\mathrm{rad}}$ RF chains, one per radiating element, and no analog or EM layer. Its digital precoder is computed by the WMMSE procedure of Inner Layer 1 and $P_\mathrm{circ} = P_\mathrm{LO} + N_t^{\mathrm{rad}}(2P_\mathrm{DAC} + P_\mathrm{RF})$. The classical hybrid AP has $N_\mathrm{RF}$ RF chains connected to $N_t^{\mathrm{rad}}$   antennas via a fully-connected phase-shifter network.  Its analog precoder is computed by Inner Layer 2 over an $N_t^{\mathrm{rad}} \times N_\mathrm{RF}$ constant-modulus manifold, its digital precoder by Inner Layer 1, and $P_\mathrm{circ} = P_\mathrm{LO} + N_\mathrm{RF}(2P_\mathrm{DAC} + P_\mathrm{RF}) + N_t^{\mathrm{rad}} P_\mathrm{PS}$ with $L_c = L_D(N_t^{\mathrm{rad}})L_\mathrm{PS}$. In the equal-power baselines, $\eta_{mk} = P_{\max}/K$, $\forall\, m,k$, and only Sub-routine 1 is executed.
 
\begin{table}[t]
	\centering \vspace{-5mm}
	\caption{Power Consumption Parameters}
	\label{tab:power_params}
	\renewcommand{\arraystretch}{1.25}
	\footnotesize
	\begin{tabular}{|l|c|c|}
		\hline
		\textbf{Parameter} & \textbf{Symbol} & \textbf{Value} \\
		\hline\hline
		PA efficiency & $\zeta_\mathrm{PA}$ & $0.27$  \\
		\hline
		DAC power & $P_\mathrm{DAC}$ & $75.8$\,mW \\
		\hline
		Local oscillator power & $P_\mathrm{LO}$ & $22.5$\,mW \\
		\hline
		RF chain power (mixer+filter+buffer) & $P_\mathrm{RF}$ & $31.6$\,mW \\
		\hline
		Phase shifter power (active) & $P_\mathrm{PS}$ & $21.6$\,mW \\
		\hline
		Varactor diode power & $P_\mathrm{var}$ & $\approx 0$ \\
		\hline
		Phase shifter insertion loss (active) & $L_\mathrm{PS}$ & $2.3$\,dB \\
		\hline
		Two-way power divider loss & $\bar{L}_D$ & $0.6$\,dB \\
		\hline
		Fixed fronthaul power per AP & $P_0$ & $825$\,mW \\
		\hline
		Fronthaul power coefficient & $\alpha_f$ & $250$\,mW/Gbps \\
		\hline
		User circuit power & $p_k$ & $200$\,mW \\
		\hline
	\end{tabular} \vspace{-5mm}
\end{table}

\begin{figure}[!t]\centering\vspace{-0mm}
	\includegraphics[width=0.4 \textwidth]{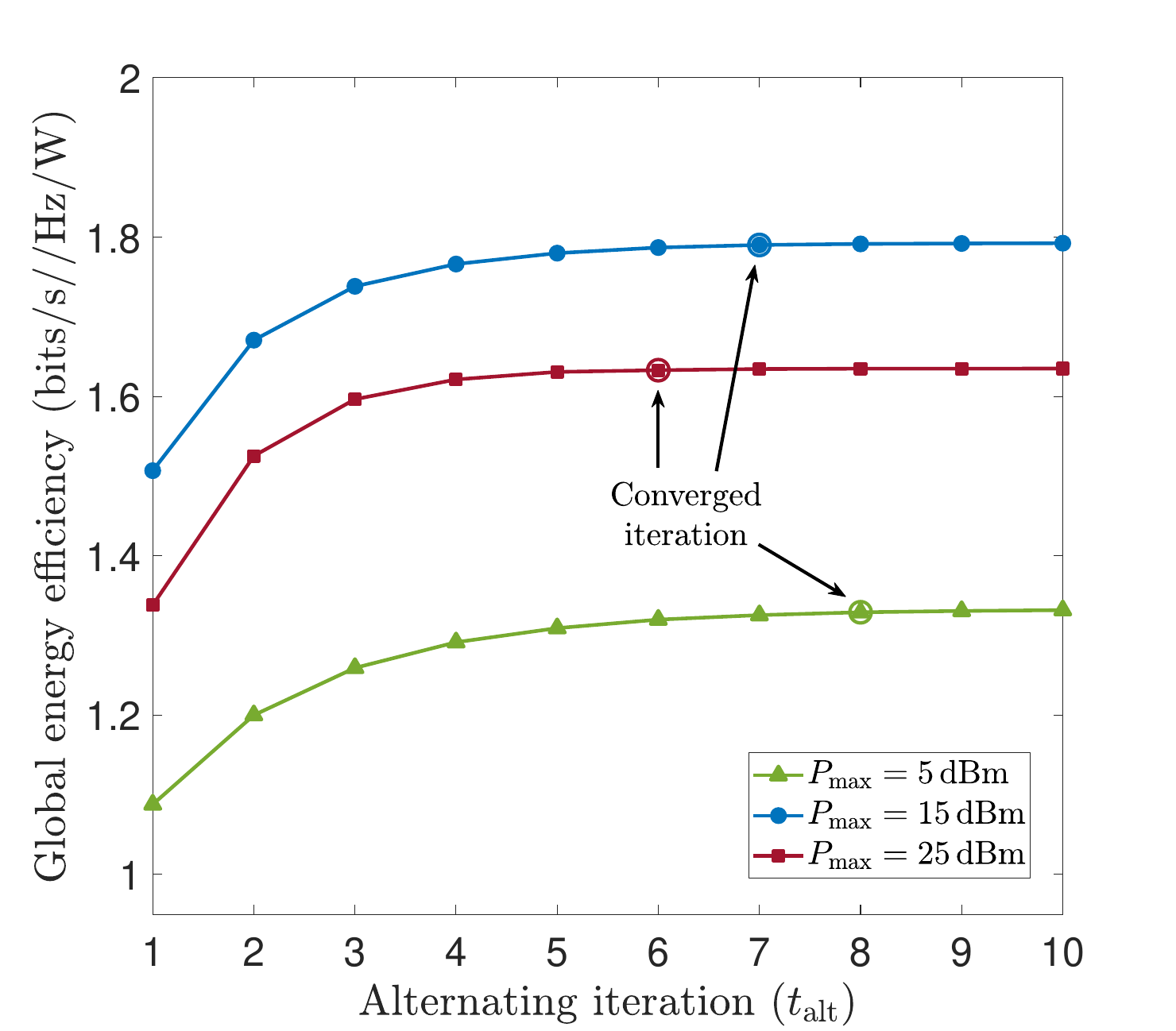}\vspace{-2mm}
	\caption{Convergence of the overall algorithm.} 
	\label{fig:convergence}\vspace{-6mm}
\end{figure}

Figure~\ref{fig:convergence} illustrates the convergence behavior of the proposed AO algorithm for three transmit power levels, $P_\mathrm{max} \in \{5, 15, 25\}$\,dBm, with $M=8$ APs, $N_\mathrm{RF}=4$, $N_e=4$, and $K=3$ users. The GEE is plotted as a function of the alternating iteration index $t_\mathrm{alt}$, initialized from the equal power allocation $\eta_{m,k} = P_\mathrm{max}/K$, $\forall\,m,k$.
It is observed that the GEE increases over the first few alternating iterations and then settles for all three power levels. The faster convergence at higher transmit power is attributed to the interference-limited operating regime, where the gradient of the GEE with respect to the power allocation coefficients is larger in magnitude, enabling the SCA subproblem to identify the optimal allocation more rapidly.


\begin{figure}[!t]\centering\vspace{-0mm}
	\includegraphics[width=0.43\textwidth]{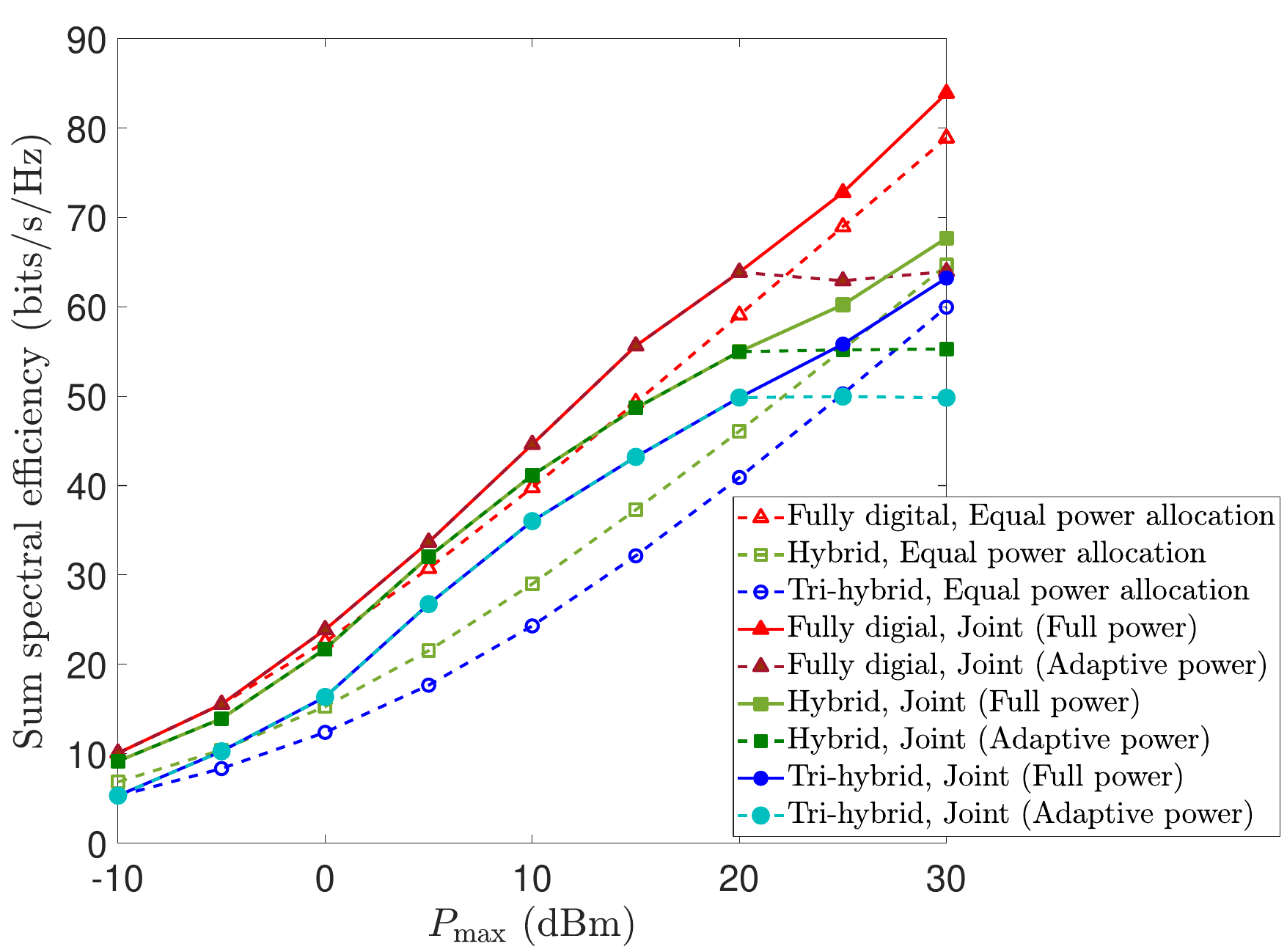}\vspace{-3mm}
	\caption{The   sum SE versus transmit power $P_{\mathrm{max}}$\,[dBm].} 
	\label{fig:rate}\vspace{-3mm}
\end{figure}

Figure.~\ref{fig:rate} plots the sum SE against the maximum per-AP transmit power ($P_\mathrm{max}$) for fully-digital, hybrid, and tri-hybrid architectures. We compare equal power allocation against our joint optimization, evaluated under both full (tight $\mathrm{C}_4$ in \eqref{eq:c4}) and adaptive (loose $\mathrm{C}_4$ in \eqref{eq:c4}) transmit power conditions. The fully-digital architecture achieves the highest SE by exploiting the full $N_t^\mathrm{rad}$-dimensional space, followed by the hybrid and tri-hybrid designs, which are limited by $N_\mathrm{RF} < N_t^\mathrm{rad}$ and analog beam steering constraints. Moreover, the proposed joint optimization with full transmit power consistently outperforms equal-power baselines, yielding SE gains of $8.11\,\%$, $19.29\,\%$, and $21.86\,\%$ at $P_\mathrm{max} = 20$\,dBm for the fully-digital, hybrid, and tri-hybrid designs, respectively. For $P_\mathrm{max} \le 20$\,dBm, the adaptive and full-power designs coincide as the power budget remains binding. Beyond $20$\,dBm, the adaptive design's SE saturates, sacrificing $23.77\,\%$, $18.33\,\%$, and $21.18\,\%$ of the full-power SE at $P_\mathrm{max}=30$\,dBm across the three architectures. This occurs because the EE objective penalizes marginal rate gains requiring excessive total power consumption. While the tri-hybrid architecture inherently underperforms fully-digital and hybrid precoders in SE, the adaptive design further trades SE for energy at high $P_\mathrm{max}$, meaning SE alone cannot characterize its performance. Hence, an investigation into the achievable  GEE  is also needed to verify the true potential of the proposed framework.


\begin{figure}[!t]\centering\vspace{-0mm}
	\includegraphics[width=0.48 \textwidth]{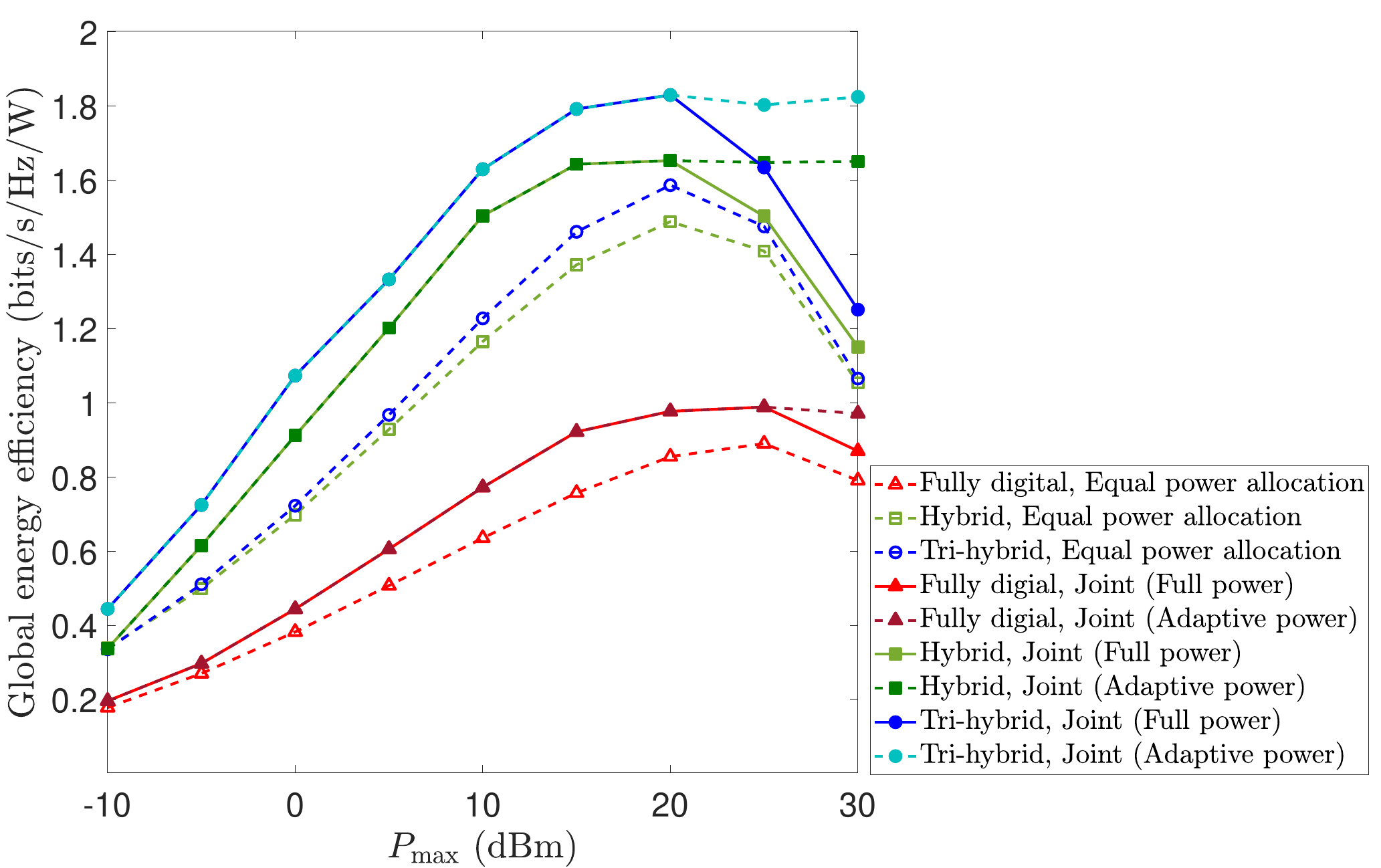}\vspace{-2mm}
	\caption{The GEE versus transmit power $P_{\mathrm{max}}$\,[dBm].} 
	\label{fig:GEE}\vspace{-7mm}
\end{figure}

Fig.~\ref{fig:GEE} plots the achievable GEE against $P_\mathrm{max}$ for the nine configurations evaluated in Fig.~\ref{fig:rate}. For the equal-power and full-power designs, the GEE exhibits a bell-shaped trend: it initially increases as SE growth outpaces fixed circuit overheads, but eventually declines at high $P_\mathrm{max}$ because SE scales logarithmically while amplifier power scales linearly. Conversely, the adaptive design's GEE coincides with the full-power design for $P_\mathrm{max}\le 20$\,dBm and saturates thereafter. By transmitting at the GEE-optimal power rather than consuming the full budget, the adaptive design avoids penalizing efficiency. Thus, at $P_\mathrm{max}=30$\,dBm, it improves GEE over the full-power design by $45.82\,\%$, $43.44\,\%$, and $11.72\,\%$ for the tri-hybrid, hybrid, and fully-digital architectures, respectively. The fully-digital architecture achieves a smaller gain due to its massive fixed circuit power, which shifts its GEE-maximizing transmit power higher.    Crucially, the proposed tri-hybrid architecture consistently outperforms all baselines across the entire $P_\mathrm{max}$ regime. This superiority stems from explicit GEE optimization and fundamentally lower circuit power via the DMA-based EM layer. While the fully-digital architecture requires 32 RF chains, both the classical hybrid and tri-hybrid architectures use only 4. Furthermore, the tri-hybrid DMA reduces the required phase-shifters from $N_t^\mathrm{rad} = 32$ to $N_t = 8$, thereby incurring lower component insertion loss than the classical hybrid precoder.

At $P_\mathrm{max} = 20$\,dBm, the proposed tri-hybrid precoder achieves 10.68\,\% and 87.32\,\% higher GEE than the classical hybrid and fully-digital precoders under joint power allocation. Furthermore, the joint optimization consistently outperforms equal-power baselines, yielding GEE gains of 14.27\,\%, 11.05\,\%, and 16.40\,\% for the fully-digital, hybrid, and tri-hybrid architectures, respectively. Ultimately, the jointly optimized tri-hybrid architecture achieves the highest GEE, and adaptive power allocation sustains this efficiency when $P_\mathrm{max}$ exceeds its optimal level, making it a compelling candidate for energy-efficient mmWave CFmMIMO systems.

In the rest of the performance evaluation, we assume that the each AP can utilize its  full transmit power if needed based on  the optimization algorithm.


\begin{figure*}[t]\vspace{-5mm}
	\centering
	\begin{subfigure}[b]{0.45\textwidth}
		\centering
		\includegraphics[width=0.99\textwidth]{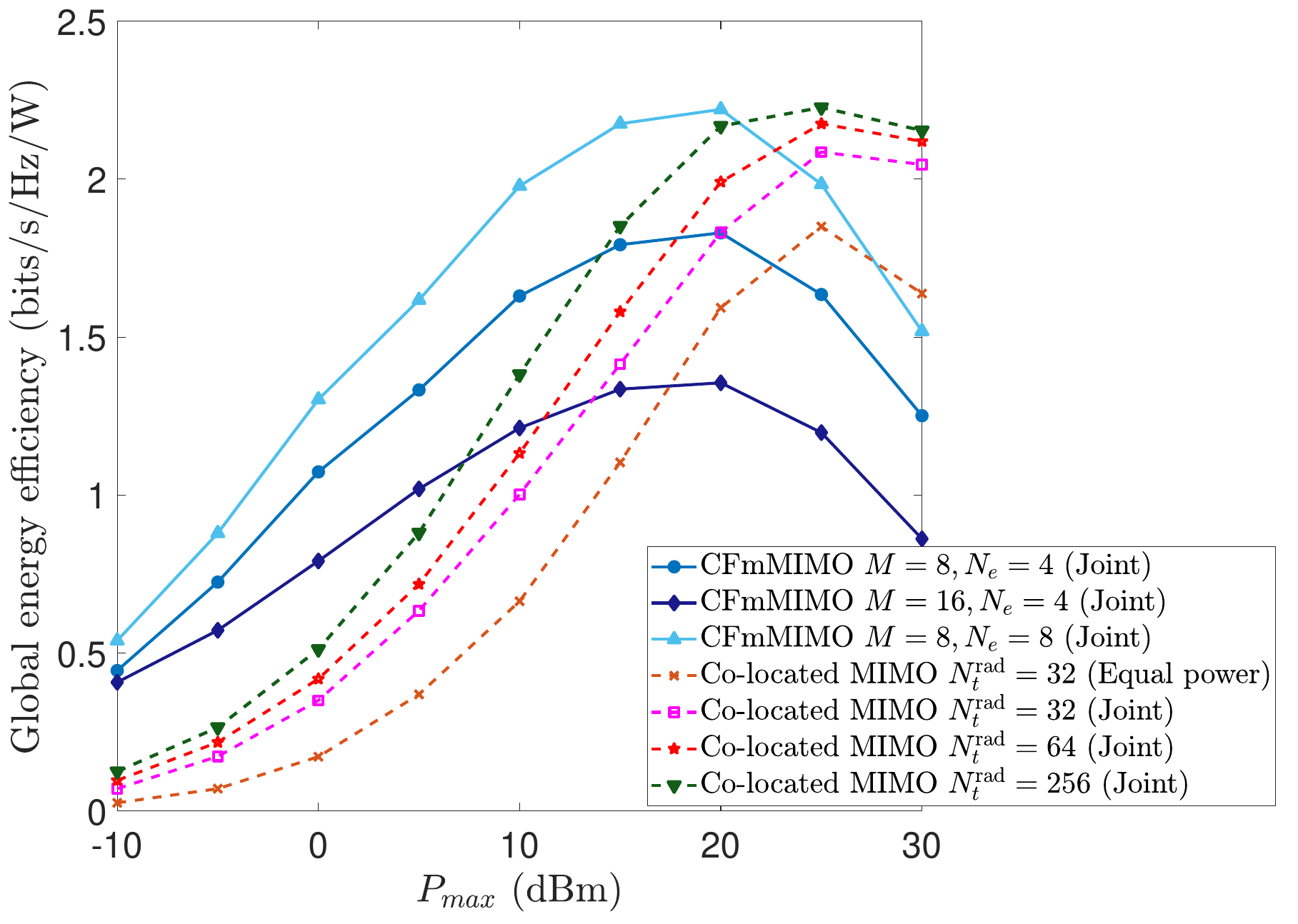}\vspace{-2mm}
		\caption{The GEE versus $P_{\mathrm{max}}$\,[dBm].}
		\label{fig:gee_vs_cl}
	\end{subfigure}
	\hfill
	\begin{subfigure}[b]{0.45\textwidth}
		\centering
		\includegraphics[width=0.85\textwidth]{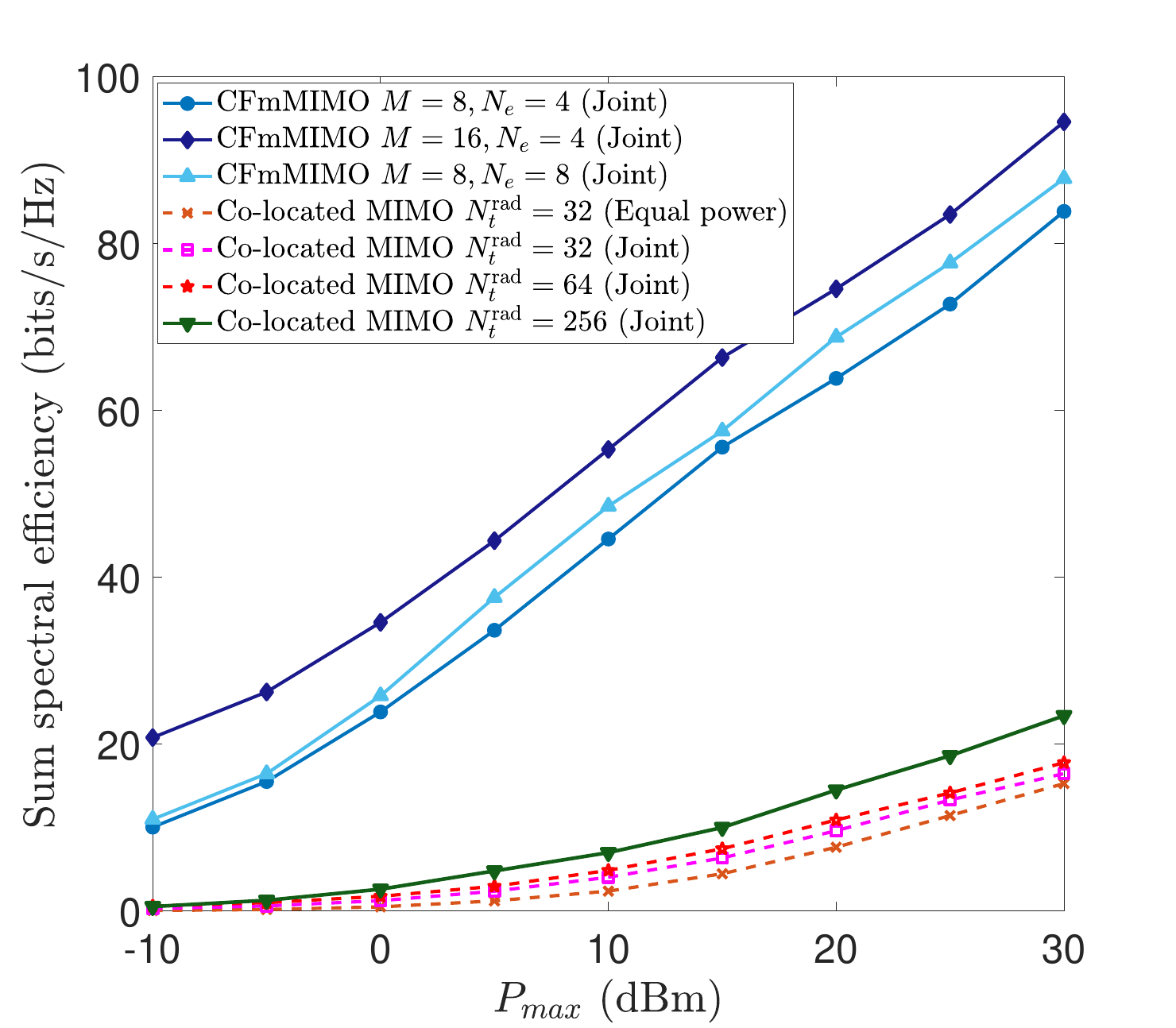}\vspace{-2mm}
		\caption{The achievable sum SE versus $P_{\mathrm{max}}$\,[dBm].}
		\label{fig:se_vs_cl}
	\end{subfigure}\vspace{-2mm}
	\caption{Comparison between Tri-Hybrid CFmMIMO and Co-located mMIMO systems.}
	\label{fig:cf_vs_cl}\vspace{-6mm}
\end{figure*}

Figures~\ref{fig:cf_vs_cl}(a) and \ref{fig:cf_vs_cl}(b), respectively, compare the GEE and SE of the tri-hybrid CFmMIMO with the proposed optimization against a co-located mMIMO baseline.
For this baseline, the distributed AP structure reduces to a single BS with $N_t^{\mathrm{rad}} \in \{32,\,64,\,256\}$, and the three-layer precoder collapses to: the EM layer operates over a single $N_t^{\mathrm{rad}}$-element DMA, the analog layer applies Riemannian manifold optimization over $N_{\mathrm{RF}}$ RF chains, and the power allocation reduces to a single-node problem with $M=1$.

As observed in Fig.~\ref{fig:cf_vs_cl}(b), the CFmMIMO consistently outperforms the co-located baseline in terms of SE across the entire useful transmit power regime. For instance, the CFmMIMO with $M=8$ and $N_{e}=4$ achieves approximately $562.5$\,\% higher SE than the co-located mMIMO with $N_t^{\mathrm{rad}}=32$ at $P_{\max}=20$\,dBm. Increasing the AP count to $M=16$ further improves the SE by approximately $16.8\%$ over $M=8$ at the same transmit power. These SE gains are attributable to the macro-diversity advantage of distributed APs in CFmMIMO, which eliminates the near-far penalty inherent to the single-cell architecture of co-located mMIMO. Even the co-located BS with $N_t^{\mathrm{rad}}=256$, which matches the total number of radiating elements of the CFmMIMO system with $M=8$ and $N_e=4$ (i.e., $M \times N_t^{\mathrm{rad}}\text{/AP} = 8 \times 32 = 256$), achieves a lower SE than CFmMIMO across the full power range, confirming that spatial distribution rather than antenna count alone is the primary driver of the SE gains.

In the comparison shown in Figs.~\ref{fig:cf_vs_cl}(a) and \ref{fig:cf_vs_cl}(b), each AP of the CFmMIMO system and the co-located BS operate under the same per-node transmit power budget $P_{\max}$. The CFmMIMO system with $M=8$ has a larger aggregate transmit power budget of $MP_{\max}$,   $MN_{\mathrm{RF}}$ RF chains and $MN_t^{\mathrm{rad}}$ radiating elements in total, compared to $N_{\mathrm{RF}}$ RF chains and $N_t^{\mathrm{rad}} \in \{32,\,64\}$ radiating elements at the co-located BS. The SE gains in Fig.~\ref{fig:cf_vs_cl}(b) thus reflect the combined benefits of  distributed transmissions   across multiple APs and   the macro-diversity gains. To isolate the contribution of spatial distribution from that of the antenna count, we also include a co-located BS with $N_t^{\mathrm{rad}}=256$, matching the total radiating elements of the CFmMIMO with $M=8$, $N_e=4$, under the same per-node power budget $P_{\max}$. The performance gap that persists between CFmMIMO, and this matched-antenna co-located baseline   quantifies the gain attributable to distributed deployment and macro-diversity, independent of the hardware.

In Fig.~\ref{fig:cf_vs_cl}(a), the GEE exhibits a distinctly non-monotone dependence on $P_{\max}$ for all configurations, peaking at a moderate power level before declining as the logarithmic saturation of the achievable rate is outpaced by the linearly increasing transmit power.
Comparing CFmMIMO configurations with $M=8$ and $M=16$ (both having $N_{e}=4$), the higher AP count yields a lower GEE despite the superior SE. This is because doubling $M$ also doubles the aggregate static circuit power $M(P_{\mathrm{circ}} + P_0)$, while the incremental rate gain diminishes at moderate-to-high power levels.
By contrast, increasing the DMA density from $N_e=4$ to $N_e=8$ while keeping $M=8$ improves the GEE by approximately $21.35\%$ at a transmit power of 20\,dBm. This is because the tri-hybrid circuit power ($P_{\mathrm{circ}}$) in \eqref{eq:Pcircth} depends only on $N_{\mathrm{RF}}$ and the number of feed ports $N_t$, but not on $N_{e}$. Consequently, the denser EM layer contributes additional passive beamforming gain at a negligible active power cost, a structural advantage unavailable to the conventional hybrid precoder architectures, whose phase-shifter count scales directly with $N_t^{\mathrm{rad}}$.
Moreover, the co-located mMIMO with $N_t^{\mathrm{rad}}=32$ and $N_t^{\mathrm{rad}}=64$ initially falls well below all the CFmMIMO curves at low-to-moderate transmit power, but crosses above the CFmMIMO with $M=8$, $N_{e}=4$ and $M=16$, $N_e=4$ at higher transmit power, achieving peak GEE values of approximately $2.09$ and $2.18$\,bits/s/Hz/W, respectively. This trend arises because the co-located BS deploys all $N_t^{\mathrm{rad}}$ antennas at a single site, enabling a high coherent array gain per RF chain ($N_t^{\mathrm{rad}}/N_{\mathrm{RF}} \in \{8,\,16\}$) that compounds with transmit power more aggressively than the per-AP gain in the cell-free case. At low transmit power, this concentration is disadvantageous due to the near-far effect. At high transmit power, the beamforming gain dominates and the single-node circuit overhead is substantially lower than $MP_{\mathrm{circ}}$ of the CFmMIMO system. The co-located BS with $N_t^{\mathrm{rad}}=256$ similarly surpasses the CFmMIMO with $M=8$, $N_e=4$ and $M=16$, $N_e=4$ at high transmit power, since the single-node circuit overhead remains fixed regardless of $N_t^{\mathrm{rad}}$, while the   array gain continues to grow with transmit power.


\begin{figure}[!t]\centering\vspace{-2mm}
	\includegraphics[width=0.4 \textwidth]{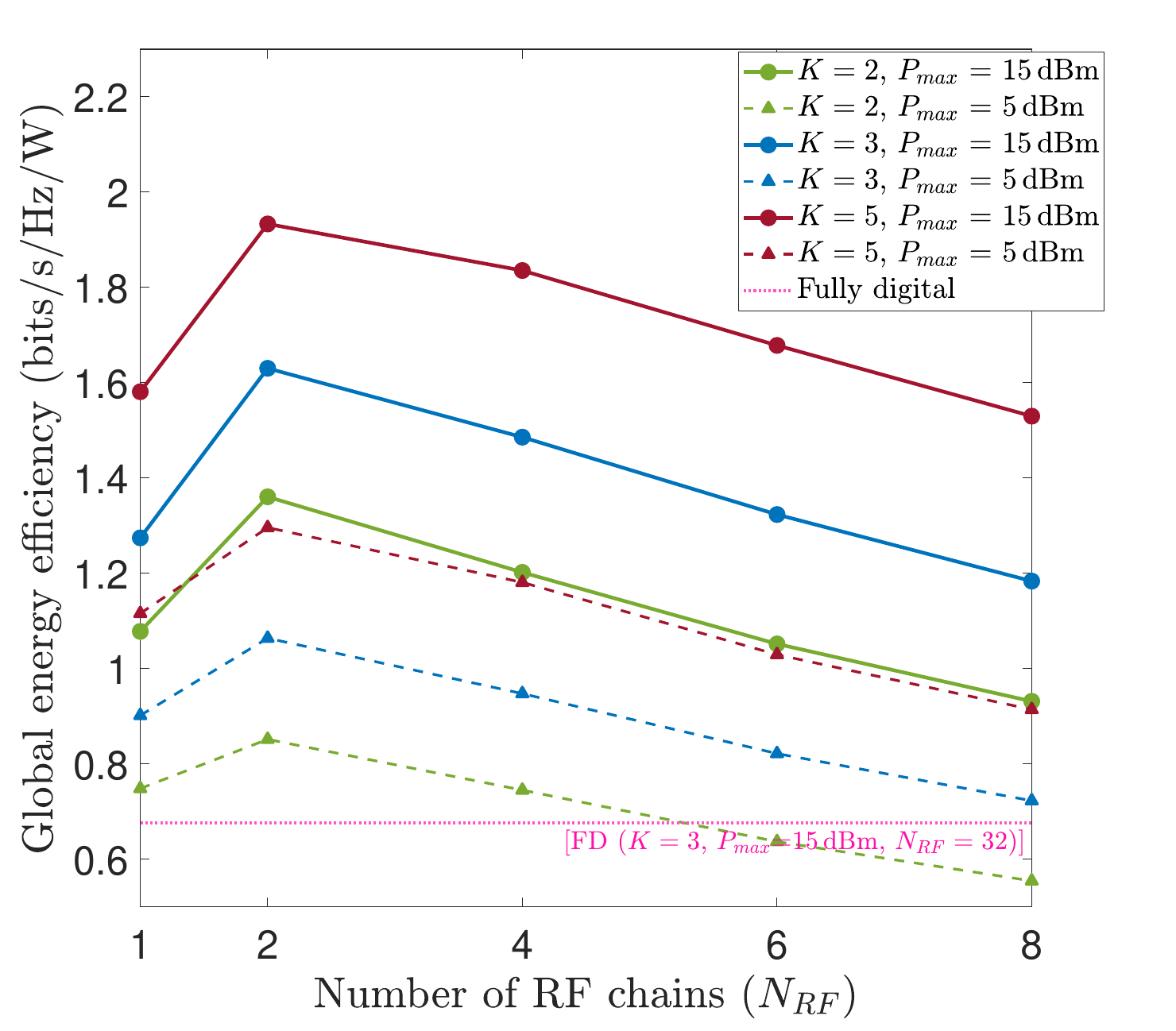}\vspace{-2mm}
	\caption{The GEE versus the number of RF chains.} 
	\label{fig:gee_vs_nrf}\vspace{-8.9mm}
\end{figure}

Figure~\ref{fig:gee_vs_nrf} plots the GEE of the proposed tri-hybrid CFmMIMO as a function of the number of RF chains $N_{\mathrm{RF}} \in \{1,2,4,6,8\}$, evaluated for $K \in \{2,3,5\}$ and $P_{\max} \in \{5,\,15\}$\,dBm. 
Since $N_s \leq N_\mathrm{RF}$ is required, $N_s = 1$ is used for $N_\mathrm{RF} = 1$, and $N_s = 2$ otherwise.
The fully-digital architecture at $K=3$ and $P_{\max}=15$\,dBm is plotted as a horizontal dotted reference/baseline. Since the fully digital precoder uses a fixed number of RF chains equal to $N_t^{\mathrm{rad}}=32$, and hence,  its GEE is a constant. Several key insights can be drawn from Fig. \ref{fig:gee_vs_nrf}.
First, the GEE of the tri-hybrid architecture is non-monotone in $N_{\mathrm{RF}}$ and peaks at $N_{\mathrm{RF}}=2$ for every combination of $K$ and $P_{\max}$.
This optimum reflects the competing effects of spatial multiplexing gain and circuit power requirements. Increasing $N_{\mathrm{RF}}$ from 1 to 2 provides a large  beamforming and multiplexing gain that outweighs the additional per-chain hardware cost $2P_{\mathrm{DAC}}+P_{\mathrm{RF}}$, since the second RF chain enables simultaneous multi-user spatial separation that a single RF chain cannot provide. Beyond $N_{\mathrm{RF}}=2$, each additional RF chain yields diminishing rate improvement while the circuit power continues to grow linearly with $N_{\mathrm{RF}}$, causing the GEE to decline. 
At $P_{\max}=15$\,dBm, the post-peak reduction resulted from  $N_{\mathrm{RF}}=2$ to
$N_{\mathrm{RF}}=8$ amounts to 31.6\,\%, 27.4\,\%, and 20.9\,\% for $K=2$, $3$,
and $5$, respectively. The progressively smaller decline with increasing $K$
indicates that a larger user count  benefits more from the the additional spatial
degrees-of-freedom offered by extra RF chains, partially compensating for their
circuit overhead.
Second, the GEE increases monotonically with the number of users $K$ at every fixed $N_{\mathrm{RF}}$ and $P_{\max}$. For example, at $N_{\mathrm{RF}}=2$ and $P_{\max}=15$\,dBm, the peak GEE values are 1.36, 1.63, and 1.93\,bits/s/Hz/W for $K=2$, $3$, and $5$, respectively, corresponding to a $42.1\%$ improvement as $K$ grows from $2$ to $5$.  This can be interpreted as follows: serving more users with the same hardware allows the fixed circuit overhead $Kp_k + MP_{\mathrm{circ}}$ to be distributed over a larger aggregate rate, thus improving the achievable GEE. 
Next, a higher $P_{\max}$ improves the GEE because the additional radiated power increases the achievable rate faster than it increases the total power consumption. At $N_{\mathrm{RF}}=2$, raising $P_{\max}$ from 5\,dBm to 15\,dBm improves the peak GEE by 59.8\,\%, 53.3\,\%, and 49.2\,\% for $K=2$, $3$, and $5$, respectively.

Figure~\ref{fig:gee_vs_nrf} also reveals that the proposed tri-hybrid precoder  outperforms the fully digital reference across all configurations. The fully digital baseline achieves a fixed GEE of   0.68\,bits/s/Hz/W. This is because  its $N_t^{\mathrm{rad}}=32$ active RF chains, each dissipating $2P_{\mathrm{DAC}}+P_{\mathrm{RF}}$ of circuit power, result in a large static power floor,  limiting its achievable GEE.  On the contrary to the transmit power, which can be  optimized jointly with the
	precoder and typically constrained to the maximum allowed limit, the circuit power consumption is a fixed 	overhead. Hence,  even when the FD precoder delivers high SE at the	moderate-to-high transmit power, the GEE remains suppressed because the rate gain is insufficient to compensate the constant circuit power consumption across all $N_t^{\mathrm{rad}} = 32$ active RF chains.

In contrast, the tri-hybrid precoding  operates with only limited,
$N_{\mathrm{RF}}=2$, active chains. It relies on the passive Lorentzian EM layer and
the unit-modulus Riemannian analog layer to perform better at very-low additional energy requirement, dramatically reducing the circuit power consumption.
Consequently, at $P_{\max}=15$\,dBm, the peak GEE gains of the tri-hybrid over the fully-digital baseline are $101.2\%$, $141.1\%$, and $185.9\%$ for $K=2$, $3$, and $5$,
respectively.
These gains are larger than the $87.32\%$ reported for Fig.~\ref{fig:GEE}, because they are evaluated at the GEE-optimal $N_\mathrm{RF} = 2$ and $P_{\max} = 15$\,dBm, whereas Fig.~\ref{fig:GEE} uses $N_\mathrm{RF} = 4$ and $P_{\max} = 20$\,dBm. The other figures use $N_\mathrm{RF} = 4$ as the nominal value, since it provides a higher SE than $N_\mathrm{RF} = 2$ while keeping the circuit power moderate.  Hence it is a balanced operating point for comparing both the SE and the GEE.


\begin{figure}[!t]\centering\vspace{-5mm}
	\includegraphics[width=0.4 \textwidth]{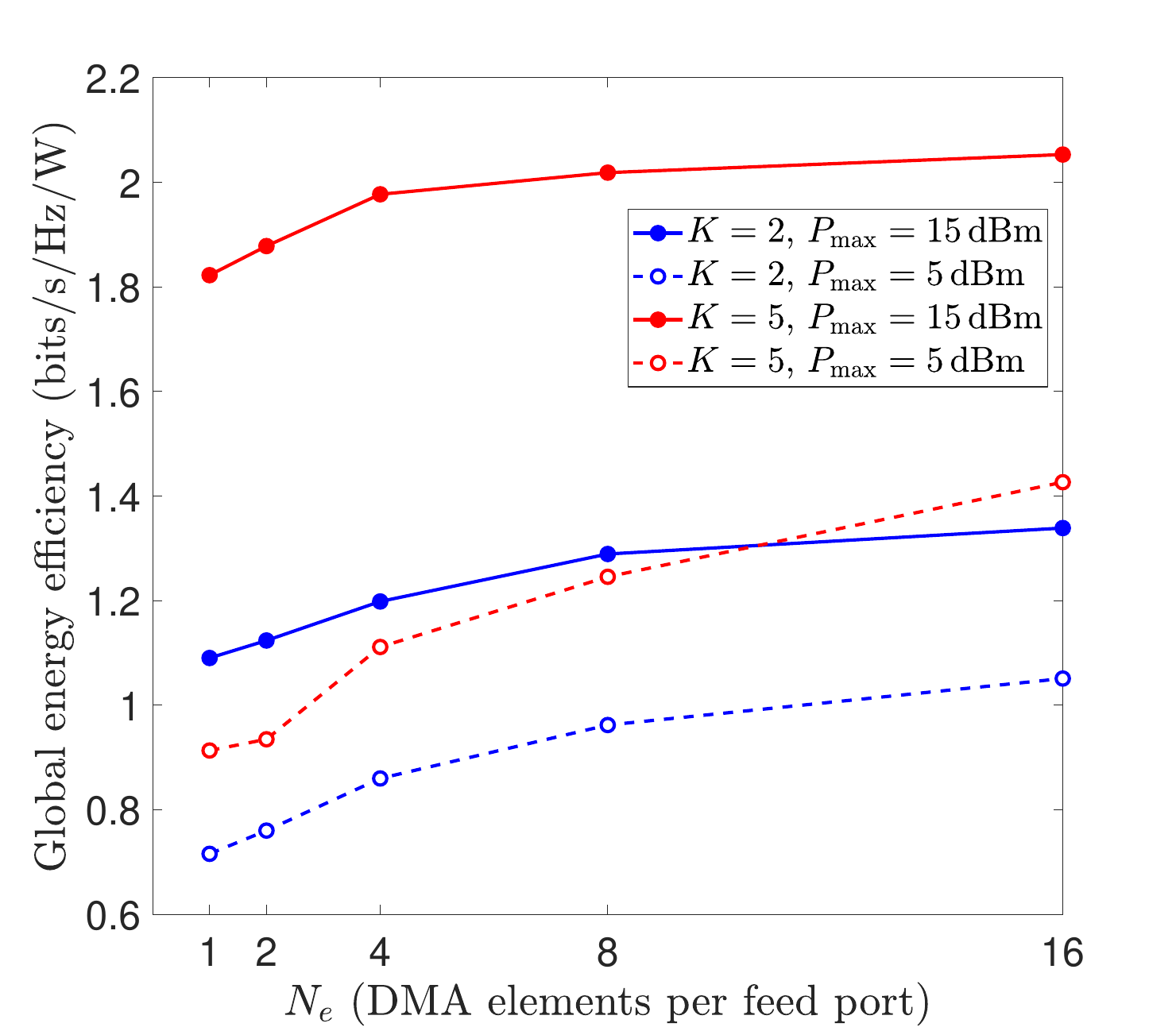}\vspace{-2mm}
	\caption{The GEE versus the number of DMA elements.} 
	\label{fig:gee_vs_ne}\vspace{-6mm}
\end{figure}

In 
Fig.  \ref{fig:gee_vs_ne},  the achievable GEE is plotted against the number of DMA elements $(N_e)$
for four cases by varying $K \in \{2,5\}$ and $P_{\max} \in \{5, 15\}$\,dBm, while keeping all other parameters fixed. In all four cases, the  GEE increases monotonically with $N_e$. 
This observation shows that  under the adopted power model, the EM layer improves the GEE without additional active power, since $P_{\mathrm{circ}}$ is independent of $N_e$ for $P_\mathrm{var} = 0$. We note two caveats. First, for PIN-diode-based DMAs with $P_\mathrm{var} > 0$, the term $N_t^{\mathrm{rad}}P_\mathrm{var}$ in \eqref{eq:Pcircth} grows with $N_e$ and reduces these gains. Second, the power model does not include the power dissipated inside the DMA. Since $|\alpha_{ip}|^2 = (1-\sin\varphi_{ip})/2$ and $\beta_{ip}$ decays along the waveguide, a fraction of the power delivered to the feed ports is not radiated, and this fraction grows with $N_e$. The gains in Fig.~\ref{fig:gee_vs_ne} should therefore be read as upper bounds for large $N_e$. 
The degree of this GEE improvement depends on the operating regime. At low $P_{\max}$ regime, the system is noise-limited, and a denser DMA array provides a larger SINR gain per added element, yielding about  $56\%$ GEE gain from $N_e=1$ to $N_e=16$ for $K=5$ at $P_{\max}=5$\,dBm. At high $P_{\max}$ regime, the system becomes interference-limited, and hence, the rate gain achieved from additional DMA elements is partially absorbed by growing inter-user interference, reducing the GEE gain to about $13\%$ for the same
$K$ at $P_{\max}=15$\,dBm. At every fixed $N_e$, $K=5$ consistently outperforms $K=2$ in terms of the GEE, since a higher user  density amortizes the fixed circuit overhead $MP_{\mathrm{circ}}$ over a higher aggregate rate. A noteworthy crossover of GEE curves occurs between the cases for $K=5$ with $P_{\max}=5$\,dBm and the $K=2$ with $P_{\max}=15$\,dBm. Here   at $N_e=8$, the former trails the latter, however, by $N_e=16$, the case with more users surpasses the case with less users by $6.5\%$, revealing that sufficiently dense DMA arrays can compensate for a reduced transmit power budget in heavily loaded networks and achieve superior GEE.

 \vspace{-1mm}
\section{Conclusion}\label{eqn:conclusion}

A tri-hybrid precoder design  for CFmMIMO systems has been investigated. 
The digital, analog, and EM domain design parameters of the three precoding layers  have been jointly optimized together with the AP transmit powers to maximize the GEE subject to   the per-AP power constraints. The transmit power optimization at each AP has also been embedded within this joint optimization as a sub-routine.  
This non-convex optimization problem has been  solved  by developing  an alternating algorithm, in which  the tri-hybrid precoder is optimized by the WMMSE method, Riemannian manifold optimization, and gradient ascent over the Lorentzian phases. The transmit  power optimization sub-routine   leverages Dinkelbach's fractional programming and an SCA-based procedure, and it can be  solved iteratively through a convex solver.  The convergence and computational complexity  of the proposed algorithm have also been evaluated. 
 Our performance evaluations and baseline comparisons  reveal that 
the proposed tri-hybrid precoder design   outperforms the fully-digital and   hybrid precoding benchmarks, while also outperforming the co-located tri-hybrid precoder deployments.
 Our design also achieves   GEE gains over an equal power allocation baseline, emphasizing  the importance of joint optimization  of cascaded precoders in  the digital, analog and EM domains, together with transmit power at each AP in CFmMIMO systems. Our SE-GEE trade-off analysis  highlights the potential of the  tri-hybrid precoder architecture as an enabler for flexible MIMO architectures as it  provides GEE gains without compromising SE gains attainable by more complicated   digital precoders.


  \linespread{1.0}

\bibliographystyle{IEEEtran}
\bibliography{IEEEabrv,References}

\end{document}